\documentclass[prb,footnoteinbib,showkeys,showpacs,twocolumn,unsortedaddress]{revtex4}
\usepackage{amsmath,amssymb,epic,eepic}
\usepackage[latin1]{inputenc}
\usepackage{graphicx}
\usepackage{subfigure}

\begin{document}

\title{Wigner function approach to single electron coherence in quantum Hall edge channels}

\author{D. Ferraro}
\author{A. Feller}
\author{A. Ghibaudo} 
\author{E. Thibierge}

\affiliation{Universit\'e de Lyon, F\'ed\'eration de Physique Andr\'e Marie Amp\`ere,\\
CNRS - Laboratoire de Physique de l'Ecole Normale Sup\'erieure de Lyon,\\
46 All\'ee d'Italie, 69364 Lyon Cedex 07, France}

\author{E. Bocquillon}
\author{G. F\`eve}

\affiliation{Laboratoire Pierre Aigrain, Département de Physique de l'Ecole Normale Supérieure, 
24 rue Lhomond, 75231 Paris Cedex 05, France}

\author{Ch. Grenier}

\affiliation{Centre de Physique Th\'eorique (CPHT), Ecole Polytechnique, 91128 Palaiseau Cedex - France}

\author{P. Degiovanni}

\affiliation{Universit\'e de Lyon, F\'ed\'eration de Physique Andr\'e Marie Amp\`ere,\\
CNRS - Laboratoire de Physique de l'Ecole Normale Sup\'erieure de Lyon,\\
46 All\'ee d'Italie, 69364 Lyon Cedex 07, France}

\begin{abstract}
Recent electron quantum optics experiments performed 
with on-demand single electron sources call for
a mixed time/frequency approach to electronic quantum coherence. 
Here, we present a Wigner function representation
of first order electronic coherence and show that is provides 
a natural visualization of the excitations emitted by recently demonstrated single
electron sources. It also gives a unified perspective on single particle
and two particle interferometry experiments. In particular, we
introduce a non-classicality criterion for single electron coherence and discuss
it in the context of Mach-Zenhder interferometry. Finally,  the electronic 
Hanbury Brown and Twiss and the Hong Ou Mandel experiments are interpreted in terms
of overlaps of Wigner function thus connecting them to signal processing. 
\end{abstract}



\maketitle


\section{Introduction}

Recent experiments have demonstrated the importance of single\cite{Ji:2003-1,Roulleau:2007-2} and two 
particle\cite{Oliver:1999-1,Henny:1999-1,Samuelsson:2004-1,Neder:2007-2} quantum coherence
in the field of quantum coherent electronics. The advent of on demand single electron 
sources\cite{Ahlers:2006-1,Feve:2007-1,Blumenthal:2007-1,Leicht:2011-1,Battista:2011-1,Hermelin:2011-1,Dubois:2013-1} 
has opened the way to a new generation of experiments dealing with
excitations having a finite spatial extension and various shapes instead 
of a continuous stream of indistinguishable
excitations. These experiments open the way
to the controlled preparation, manipulation and characterization of single to few electron
excitations in ballistic conductors, an emerging field 
called electron quantum optics\cite{Degio:2011-1}. Since these experiments
access time scales comparable to the coherence time of electrons within conductors 
through finite frequency current\cite{Gabelli:2006-1,Bocquillon:2012-2} and noise
measurements\cite{Mahe:2010-1,Parmentier:2010-1}, a time resolved 
approach to electronic quantum coherence is required.

\medskip

In this paper, we present a unified view of the various representations of single electron
coherence in quantum Hall edge channels and we introduce in the present context 
a time/frequency representation
based on the Wigner function introduced in quantum mechanics\cite{Wigner:1932-1}  and signal 
processing\cite{Ville:1948-1}. The Wigner function is commonly used in quantum 
optics\cite{book:Walls-Milburn} and has been recently been measured in cavity QED
to demonstrate the decay of quantum superpositions of two quasi-classical states
of the electromagnetic field\cite{Deleglise:2008-1}.

Although all the representations of electronic coherence contain exactly the same
information, each of them has its advantages and drawbacks. 
First, the time domain representation of single electron coherence is suitable for
analyzing time-dependent aspects as well as to define the proper notions 
of coherence and dephasing times\cite{Haack:2012-2}.
However, information on the electron or hole 
nature of excitations in hidden in the phase of this quantity. 
On the other hand, the frequency domain representation
is perfect for discussing the nature of excitations and is the natural representation for the 
electronic analogue of homodyne tomography\cite{Degio:2010-4} but it is not well suited 
for describing real time aspects. 
The mixed time/frequency
representation called the electronic Wigner function combines the advantages 
of both representations:
it gives a direct access to both the time evolution and energy content of single 
electron coherence. Moreover,
it provides a natural non-classicality criterion for single electron coherence and 
for example, enable us to discuss non-classicality in Mach-Zenhder interferometry.
 
Historically, the Wigner function has been introduced in the theory of quantum transport at
the end of the 80s to understand the limits of a semi-classical 
treatment in semiconductors\cite{Jacoboni:2004-1}, to study phonon interaction
effects\cite{Bordone:1999-1} and also to model various quantum  
devices built from semiconductors\cite{Frensley:1987-1,Kluksdahl:1989-1}. 
Although we deal with the same concept, we consider here low dimensional conductors 
in which many-body effects as well as interaction and decoherence
in the presence of the Fermi sea are crucial. Moreover, our primary
motivation is to discuss the various representation of single electron coherence
and show the relevance of the electronic Wigner function for understanding electron quantum optics
experiments.
  
\medskip

The price to pay is that the electronic Wigner function, as of today, cannot 
be measured directly at a given time and frequency 
contrary to cavity QED experiment\cite{Lutterbach:1997-1,Melo:2006-1}. Nevertheless, 
we will show that it provides a unified view of single electron interference experiments as well as of 
two particle interference experiments based on the
Hanbury Brown and Twiss\cite{Hanbury:1956-2} (HBT) effect. These include the 
Hong, Ou, Mandel (HOM) experiment\cite{Hong:1987-1} which involve two electron
interferences from two different sources in full generality.
By doing so, we will greatly simplify the discussion of the various protocols for reconstructing
single electron coherence. We show that the recently proposed 
tomography protocol\cite{Degio:2010-4} does indeed
directly reconstructs the electronic Wigner function. Moreover, this approach, 
by providing a natural visualization of 
single electron coherence, suggests that specifically designed alternative tomography 
protocols may lead to single electron reconstruction
from less measurements than our generic homodyne protocol. We finally sketch
a possible connection between the search for such optimized tomography 
protocols and the problematic of compressed sensing in 
signal processing\cite{Book:Compressed-Sensing}.

\medskip

This paper is structured as follows: in Sec.~\ref{sec:eqo:G1}, single electron coherence 
and its relation to Glauber's coherence in quantum optics is revisited. 
We then discuss the time and frequency representations
of this quantity and we introduce the electronic Wigner function. 
Section~\ref{sec:examples} is devoted to examples: we first consider
single electron coherence in the presence of a classical voltage drive. 
The cases of a sinusoidal drive and of Lorentzian pulses\cite{Dubois:2013-1,Dubois:PhD} are discussed.  
We then discuss the electronic Wigner function emitted
by the mesoscopic capacitor used as an on demand single electron source\cite{Feve:2007-1}.
Section \ref{sec:interferometry} is devoted to interferometry experiments, starting 
with Mach-Zehnder interferometry and
then discussing the HBT and HOM two particle interference experiments.

\section{Single electron coherence}
\label{sec:eqo:G1}

\subsection{Definition and simple examples} 
\label{sec:eqo:G1:basics}

Mach-Zehnder interferometry has shown the importance of first order electron coherence.
It is usually defined as the non equilibrium Keldysh Green's function already used to describe
electronic coherence in the many body approaches to the decoherence problem 
in diffusive conductors\cite{Golubev:1999-3}:
\begin{equation}
\label{eq:SPC:1-e}
\mathcal{G}^{(e)}_{\rho}(x,t;y,t') = \mathrm{Tr}(\psi(x,t)\rho\,\psi^\dagger(y,t'))\,.
\end{equation}
As noticed by Glauber {\it et al}\cite{ Cahill:1999-1}, this 
correlator is also relevant for atom-counting experiment with fermionic cold atom system.
A similar quantity can be defined for hole excitations:
\begin{equation}
\mathcal{G}^{(h)}_{\rho}(x,t;y,t') = \mathrm{Tr}(\psi^\dagger(x,t)\rho\,\psi(y,t'))\,.
\label{eq:SPC:1-h}
\end{equation}
The electron and hole coherences at coinciding times are related using 
the canonical anti commutation relations
\begin{equation}
\label{eq:SPC:eh}
\mathcal{G}^{(e)}(x,t;y,t)=\delta(x-y)-\mathcal{G}^{(h)}(y,t;x,t)\,.
\end{equation}
The first important difference with photon quantum optics comes 
from the fact that, in a metallic conductor, the Fermi sea, which plays here the role of the vacuum state,
has a non vanishing single particle coherence
whereas coherence vanishes in the photon vacuum. At zero temperature, 
the single electron coherence within a single chiral 
channel at equal times is given by:
\begin{equation}
\label{eq:SPC:2}
\mathcal{G}^{(e)}_{\mu}(x,t;y,t) = \frac{i}{2\pi }\,\frac{e^{ik_F(\mu)(x-y)}}{y-x+i0^+}
\end{equation}
where $k_{F}(\mu)$ denotes the Fermi momentum associated with the chemical potential 
$\mu$ of the edge channel under consideration.
At non-vanishing electronic temperature $T_{\mathrm{el}}$, these 
correlators decay over the thermal length scale $l(T_{\mathrm{el}})=\hbar v_{F} /k_BT_{\mathrm{el}}$ 
where $v_{F}$ denotes the Fermi velocity:
\begin{equation}
\label{eq:SPC:3}
\mathcal{G}^{(e)}_{\mu,T_{\mathrm{el}}}(x,t;y,t) = 
\frac{-i}{2l(T_{\mathrm{el}})}\,
\frac{e^{ik_F(\mu)(x-y)}}{\sinh{\left(\frac{\pi(y-x)+i0^+}{l(T_{\mathrm{el}})}\right)}}
\end{equation}
This suggests to decompose the single electron coherence into 
a Fermi sea contribution due  to the chemical potential $\mu$ of
the conductor and a contribution due to excitations above this 
ground state\cite{Degio:2010-4,Haack:2012-2}:
\begin{equation}
\label{eq:SPC:3}
\mathcal{G}^{(e)}_{\rho}(x,t;y,t')=\mathcal{G}^{(e)}_{\mu,T_{\mathrm{el}}}(x,t;y,t')
+\Delta\mathcal{G}_{\rho}^{(e)}(x,t;y,t')\,.
\end{equation}
The case of an ideal single electron excitation helps clarifying the physical meaning of
$\Delta\mathcal{G}^{(e)}_{\rho}$. Such a source generates a many body state 
obtained from a Fermi sea by adding one extra-particle in a normalized 
wave packet $\varphi_{e}$:
\begin{equation}
\label{eq:SPC:4}
\psi^\dagger[\varphi_{e}]|F\rangle = \int_{-\infty}^{+\infty} \varphi_{e}(x)\,
\psi^\dagger (x)\,|F\rangle\,dx
\end{equation}
where $|F\rangle $ denotes the Fermi sea at a fixed chemical potential. In
momentum space, $\varphi_e$ only has components on single particle 
states above the Fermi level. Then, Wick's theorem
leads to single electron coherence at initial time:
\begin{equation}
\label{eq:SPC:5}
\Delta\mathcal{G}^{(e)}_{\psi^\dagger[\varphi_{e}]|F\rangle}(x,0;y,0)=\varphi_{e}(x)\,\varphi^*_{e}(y)\,.
\end{equation}
In the same way, the single electron coherence of 
the state obtained by adding a single hole excitation to the Fermi sea
$\psi[\varphi_{h}]|F\rangle$ ($\varphi_{h}(k)=0$ above the Fermi level) is given by
$\Delta\mathcal{G}^{(e)}_{\psi[\varphi_{h}]\,|F\rangle}(x,0;y,0)=
-\varphi_{h}(x)\,\varphi_{h}^*(y)$. The $-$ sign reflects the fact that a 
hole is the absence of an electron in the Fermi sea. 

\medskip

The excess single electron coherence $\Delta \mathcal{G}^{(e)}_{\rho}(x,t;y,t)$ contains 
information on both the shape of the wave packet
and its phase dependence. More precisely, $\Delta\mathcal{G}^{(e)}_{\rho}(x,t;x,t)$ encodes the average
density and thus the shape of the wave packet whereas the $x-y$ dependence encodes the
phase dependence. The shape of the wave packet gives access to its length (or duration in the
time domain) which we denote by $l_1$ (resp. $T_1$).

But a realistic single electron source does not necessarily emit a perfectly coherent
wave packet: it can also emit a statistical mixture of them or the electrons may have experienced some
decoherence due to Coulomb interactions\cite{Bocquillon:2013-1}. 

How can we measure the decay of single electron coherence? As 
recently noticed by G.~Haack {\it et al}\cite{Haack:2012-2}, 
the first degree of coherence originally introduced by Mandel
for photons\footnote{Working with equal time coherences, we drop the time dependence
for simplicity. Note that in the present case, $g^{(1)}(x,l)$ is well defined only when the product 
$\Delta\mathcal{G}^{(e)}_{\rho}(x+l/2,x+l/2)\,\Delta\mathcal{G}^{(e)}_{\rho}(x-l/2,x-l/2)$
is positive. But for a source emitting only electrons, this is always the case.}:
\begin{equation}
\label{eq:SPC:g1}
g^{(1)}(x,l)=
\frac{\Delta\mathcal{G}^{(e)}_{\rho}(x+\frac{l}{2},x-\frac{l}{2})}{\sqrt{\Delta\mathcal{G}^{(e)}_{\rho}(x+\frac{l}{2},x+\frac{l}{2})
\,\Delta\mathcal{G}^{(e)}_{\rho}(x-\frac{l}{2},x-\frac{l}{2})}}
\end{equation}
A very simple description of decoherence consists in introducing
a phenomenological decoherence coefficient in front of the excess single electron coherence
associated with a coherent wave packet\cite{Bocquillon:2013-1}:
\begin{equation}
\label{eq:SPC:decoherence}
\Delta \mathcal{G}^{(e)}(x,y)\simeq\mathcal{D}(x-y)\,\varphi_{e}(x)\varphi_{e}(y)^*
\end{equation}
In this case, the decaying length of this decoherence coefficient is precisely $l_{\phi}$. However
such an approach to decoherence is justified only when the electron under consideration 
can still be distinguished from electron/hole
pairs generated in the Fermi sea by Coulomb interactions. Such a form of
decoherence can then obtained in a single electron model under the influence of an
harmonic environment\cite{Degio25}. 
Otherwise, a more complete approach should
be used\cite{Degio:2009-1}. Nevertheless, using Eq.~\eqref{eq:SPC:decoherence} into
Eq.~\eqref{eq:SPC:g1} shows that $g^{(1)}(x,l)$ decays over the same length scale $l_{\phi}$
as $\mathcal{D}(x-y)$ which is called the dephasing length (the 
dephasing time in the time domain). 

It then follows from \eqref{eq:SPC:g1} that 
the length scale governing the decay of the single electron 
coherence $\Delta \mathcal{G}^{(e)}_\rho (x,y)$ combines the decaying 
length of the wave packet and $l_{\phi}$:
\begin{equation}
\frac{1}{l_2}=\frac{1}{2l_1}+\frac{1}{l_\phi}\,.
\end{equation}
This formula mimics the famous expression of the total decoherence rate in NMR in terms of
the relaxation and dephasing rates\cite{Haack:these,Haack:2012-2}. 
 
\subsection{Analogy with Glauber's coherences}
\label{sec:eqo:G1:analogies}

In Glauber's approach to photo-detection\cite{Glauber:1962-1}, 
a photon detector is a quantum device designed to detect a single photon. 
In such a detector, a single photon causes the photoionization of an atom and the emitted 
electron is then amplified to give a macroscopic signal. 
Old photomultipliers work exactly this way: the incoming photon
is absorbed, leading to the emission of a single electron which is then amplified by the 
secondary emission of electrons in dynodes. The initial photo-emission stage can be described
using elementary time dependent perturbation theory. The resulting photo-detection signal is then
obtained as
\begin{equation}
I_{D}(t)=\int_{0}^t K_{D}(\tau,\tau')\, 
\mathcal{G}^{(1)}_{\rho}(x_D,\tau;x_{D},\tau')
\,d\tau d\tau'
\end{equation}
where $x_{D}$ denotes the position of the detector. The quantity 
$\mathcal{G}^{(1)}_{\rho}(x_D,\tau|x_{D},\tau')$ only depends on the quantum state of the
electromagnetic field and is indeed Glauber's single photon coherence 
function\cite{Glauber:1963-1}:
\begin{equation}
\label{eq:ego:G1-photons}
\mathcal{G}^{(1)}_{\rho}(x,\tau;x',\tau')=\mathrm{Tr}\left(E^{(+)}(x,\tau)\,\ldotp 
\rho\ldotp E^{(-)}(x,\tau')\right)
\end{equation}
where $E^{(\pm)}$ denote the positive (resp. negative) frequency part of the electric field
operator and $\rho$ is the electromagnetic field initial density operator. The function
$K_{D}(\tau,\tau')$ characterizes the response of the detector to the absorption of a single photon.
Broadband detectors have a local response in time 
$K_{D}(\tau-\tau')\simeq \delta(\tau-\tau')$ and therefore 
measure the (integrated) photocount: 
$I_{D}(t)\simeq \int_0^{t} \mathcal{G}^{(1)}_{\rho}(x_{D},t';x_{D},t')\,dt'$. On the other hand,
narrow band detectors select a single frequency and therefore 
measure the Fourier transform of Glauber's single photon
coherence with respect to time.

\medskip

The analogy between Glauber's single photon coherence in Eq.~\eqref{eq:ego:G1-photons} and the single
electron coherence function given by Eq.~\eqref{eq:SPC:1-e} is then obvious. 
But what would be the analogous of photo-detection for electrons? 
The idea is simply to extract an electron from the conductor we want to probe and to
amplify the corresponding charge deposited into the detector. 
Naturally, the stage corresponding to photo-ionization is
simply tunneling of electrons from the conductor into the 
detector which could for example be an STM tip
or a nearby dot. Of course this approach does not
take into account the electrostatic coupling between the conductor and the detector. 
Assuming a pointlike detector located at position $x$, 
the average tunneling current from the conductor to the detector contains two 
contributions arising from electron transmitted from
the conductor to the reservoir and vice versa:
\begin{eqnarray}
I_{D}(t) & = & \int_{0}^t \left(\mathcal{G}_{\rho}^{(e)}(x,\tau;x,\tau')K_{a}(\tau,\tau') \right. \nonumber \\
& - & \left. \mathcal{G}_{\rho}^{(h)}(x,\tau;x,\tau')K_{e}(\tau,\tau')\right)\,d\tau d\tau'
\end{eqnarray}
In this expression, $K_{a}$ and $K_{e}$ characterize the detector and respectively account for 
available single electron and hole states within the reservoir and for the eventual energy filtering of the detector. 
Such a detection scheme has been recently implemented experimentally to study electron relaxation in 
quantum Hall edge channels\cite{Altimiras:2010-1}: in these experiments, a quantum dot is used to filter energies. 
This corresponds to a narrow-band detection in the quantum optics language. 

\subsection{Representations of single electron coherence}
\label{sec:eqo:G1:representations}

\subsubsection{The time and frequency domains}
\label{sec:eqo:G1:time-frequency}

Since measurements are usually performed locally, 
let us consider the single electron coherence function at a given position $x$:  
$\mathcal{G}^{(e)}_{\rho,x}(t,t')=\mathcal{G}^{(e)}_{\rho}(x,t;x,t')$. In the time domain, 
the diagonal $\mathcal{G}^{(e)}_{\rho,x}(t,t)$
is nothing but the average electronic density
at time $t$ and position $x$. Consequently, in chiral edge channels with Fermi velocity $v_{F}$, the excess current with
respect to the chemical potential $\mu$ is 
\begin{equation}
\langle i(x,t)\rangle_{\rho} =-ev_{F}\Delta\mathcal{G}^{(e)}_{\rho,x}(t,t)\,.
\end{equation}
The off diagonal ($t\neq t'$) excess single electron coherence $\Delta\mathcal{G}^{(e)}_{\rho,x}(t,t')$ is complex. 
Introducing $\bar{t}=(t+t')/2$ and $\tau=t-t'$, the decay of $|\Delta\mathcal{G}^{(e)}_{\rho,x}(t,t')|$ with increasing $\tau$
defines the coherence time of the source at time $t$. So this representation is indeed appropriate to discuss the coherence
time as well as to discuss time dependence. But it is not well suited for understanding 
the nature (electron or hole) of the excitations emitted by the source since it is encoded in the $t-t'$ dependence of
the phase of  $\Delta\mathcal{G}^{(e)}_{\rho,x}(t,t')$.

\medskip

\begin{figure}
\includegraphics[width=8cm]{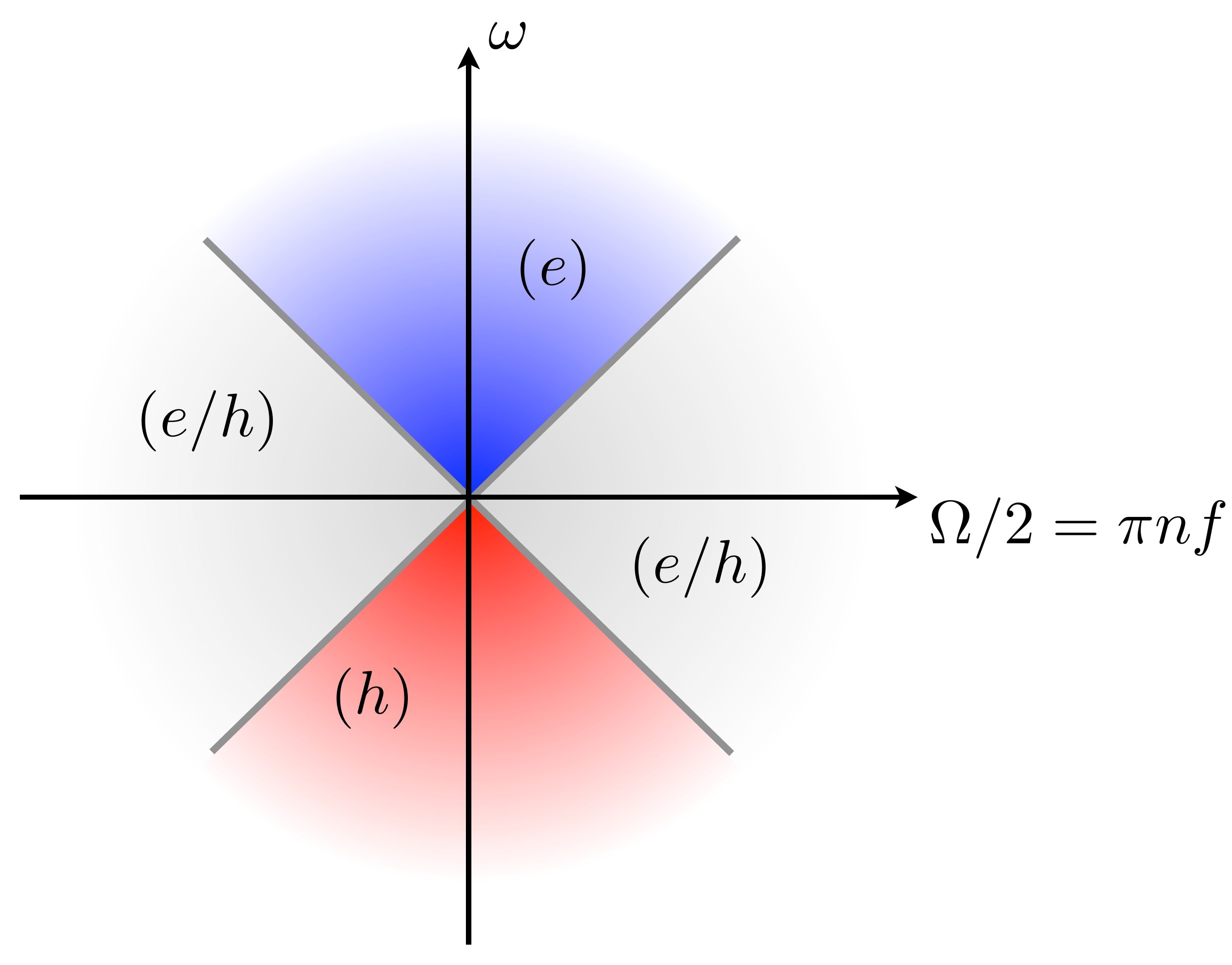}
\caption{\label{fig:quadrants} (Color online) 
Decomposition of the Fourier plane $(\omega,\Omega)$ in four quadrants. 
The (e) quadrant (in blue) represents
the contribution of purely electronic excitations; the (h) quadrant (in red) represents 
the contribution of hole excitations. The two (e/h) quadrant (off diagonal in grey) 
encode the contribution of electron/hole coherences. The diagonal ($\Omega=0$, 
vertical axis) gives the electron occupation number. By restricting the single electron
coherence to the (e) (resp. (h)) quadrants and normalizing them properly provides an 
effective density operator for electron (resp. hole) excitations. For a periodically 
driven source with period $T=1/f$, $\Omega$ is a multiple of $2\pi nf$.}
\end{figure}

Going to the frequency domain gives access to the nature of excitations. The single electron
coherence in the frequency domain is defined as the double Fourier transform:
\begin{equation}
\widetilde{\mathcal{G}}^{(e)}_{\rho,x}(\omega_{+},\omega_{-})=
\int \mathcal{G}^{(e)}_{\rho,x}(t,t')\,e^{i(\omega_{+}t-\omega_{-}t')}\,dt\,dt'\,.
\end{equation}
As shown on Fig.~\ref{fig:quadrants}, the Fourier plane can then be divided into four quadrants. 
The (e) or electron quadrant defined by $\omega_{+}>0$ and $\omega_{-}>0$ contains the contribution
of excitations that correspond to single particle levels having positive 
energies with respect to the chemical potential $\mu=0$. 
The (h) or hole quadrant defined by $\omega_{+}<0$ and $\omega_{-}<0$ contains the contribution
of excitations that correspond to single particle levels with 
negative energies with respect to the chemical potential $\mu=0$. Finally, the (e/h) or electron/hole quadrants
 are defined by $\omega_{+}\omega_{-}\leq 0$ correspond to the coherence between electronic
and hole excitations. These coherences 
are produced in the presence of a superposition of electron and hole excitations\cite{Degio:2010-4}.
At this point, it is worth mentioning that the electronic or hole nature of excitations is defined with respect to
a given chemical potential, here chosen as $\mu=0$. With respect to another chemical potential, excitations
would be categorized differently. 

Using the Fourier decomposition of the single
electron creation and destruction operators, the single electron coherence in the frequency domain 
satisfies
\begin{equation}
\label{eq:SPC:frequency-domain}
\frac{v_F}{2\pi}\,\widetilde{\mathcal{G}}^{(e)}_{\rho,x}(\omega_+,\omega_-)
=\langle c^\dagger(\omega_-)\,c(\omega_+)\rangle_\rho\,
\end{equation}
where $c(\omega)$ and $c^{\dagger}(\omega)$ respectively denote the fermion destruction and creation
operators at position $x$ (see Appendix. \ref{appendix:coherence}).
The electron occupation number can thus be recovered from the diagonal in the frequency domain. 
In particular, an equilibrium state corresponds to a singular single electron coherence in the
frequency domain:
\begin{equation}
\label{eq:SPC:Fermi-sea:frequency}
\widetilde{\mathcal{G}}^{(e)}_{\mu,T_{\mathrm{el}}}(\omega_{+},\omega_{-})=
\frac{2\pi}{v_F}\delta(\omega_{+}-\omega_{-})
f_{\mu,T_{\mathrm{el}}}\left(\frac{\omega_{+}+\omega_{-}}{2}\right).
\end{equation}
A convenient 
way to visualize the $(\omega_+,\omega_-)$ plane\cite{Degio:2010-4} uses $\omega=(\omega_++\omega_-)/2$ and
$\Omega=\omega_+-\omega_-$ which are respectively conjugated to $t-t'$ and $(t+t')/2$. 
Figure \ref{fig:quadrants} depicts the (e), (h) and (e/h) quadrants with respect to $\mu=0$ in these coordinates.

The fact that, at zero temperature, the Fermi sea coherence \eqref{eq:SPC:Fermi-sea:frequency} is localized in the frequency domain
leads to Cauchy-Schwarz inequalities discussed in Appendix \ref{appendix:coherence}. They imply that,
at zero temperature, a source that does not emit any excess hole excitation (resp. electronic excitation)
has non vanishing single electron coherence only within the (e) (resp. (h)) quadrant. 

\medskip

To summarize, the frequency domain representation of the single electron 
coherence is clearly well suited to visualizing the nature of excitations with 
respect to a given chemical potential. But recovering time dependence
aspect is more difficult since it is encoded in the phase dependance of the off diagonal terms
$\widetilde{\mathcal{G}}^{(e)}_{\rho,x}(\omega_{+},\omega_{-})$ for $\omega_{+}\neq \omega_{-}$.
Since it is difficult to detect a single electron in one shot on a sub nanosecond time scale,
it may be easier to access single electron coherence in the frequency domain
than in the time domain\cite{Degio:2010-4} although a protocol has recently been proposed
using Mach-Zehnder interferometry\cite{Haack:2011-1}. We shall come back to this question in 
section \ref{sec:interferometry}. 

However, working with sources that are able to inject a single to few 
electrons and holes strongly rises the need for a representation of the single 
electron coherence giving access to real time phenomenon as well as 
to the nature of excitations. Such a time/frequency representation has been 
known for a long time in quantum statistical physics: it is the Wigner 
function\cite{Wigner:1932-1,Ville:1948-1} which we will now discuss for single 
electron coherence.

\subsubsection{The Wigner function}
\label{sec:eqo:G1:Wigner}

In order to capture both the $\bar{t}=(t+t')/2$ dependence of 
single electron coherence and the nature of excitations, we define 
the Wigner distribution function as:
\begin{equation}
\label{eq:Wigner:definition}
W_{\rho,x}^{(e)}(\bar{t},\omega)=\int v_F\mathcal{G}_{\rho,x}^{(e)}
(\bar{t}+\tau/2,\bar{t}-\tau/2)\,e^{i\omega \tau}d\tau\,
\end{equation}
which is dimensionless due to the presence of the velocity $v_F$. 
In a similar way, one would define the Wigner function for the hole excitations 
by substituting $\mathcal{G}_{\rho,x}^{(h)}$ into \eqref{eq:Wigner:definition}. Note that due to the
hermiticity properties \eqref{eq:coherences:hermiticity} 
of the single electron and single hole coherences, these Wigner functions are real. 

\medskip

Originally, the Wigner function was introduced to provide a bridge between classical and quantum mechanics
for a single particle\cite{Wigner:1932-1}. In classical mechanics, a particle has a well defined position and 
momentum and an ensemble of such
particles is represented by a probability distribution over phase space. For a quantum particle, the uncertainty
principle prevents the particle to be perfectly localized. Still, the Wigner distribution function is the proper
generalization of the probability distribution: it is a real distribution over phase space whose 
integration over position (resp. momentum) gives the probability distribution of the momentum (resp. position).
The Wigner function is thus normalized but contrarily to the classical probability distribution, it is not always positive. 

\medskip

Significant differences are expected
for the Wigner function associated with single electron coherence. In the stationary case, 
the Wigner function defined by Eq.~\eqref{eq:Wigner:definition} 
is nothing but the time independent electronic distribution function at position $x$:
\begin{equation}
\label{eq:Wigner:stationary}
W_{\rho,x}^{(e)}(\bar{t},\omega)=f_{e}(\omega,x)\,.
\end{equation}
For a Wigner function to be interpreted as a time dependent electronic distribution function, it has 
to satisfy $0\leq W_{\rho,x}^{(e)}(\bar{t},\omega)\leq 1$. The positivity condition is needed in order 
to have an interpretation as a probability density. In a chiral system, 
the electronic and hole Wigner distributions are related by
\begin{equation}
\label{eq:Wigner:e-h-relation}
W_{\rho,x}^{(h)}(\bar{t},\omega)=1-W^{(e)}_{\rho,x}(\bar{t},-\omega)\,
\end{equation}
where the minus sign in front of $\omega$ in the r.h.s 
reflects the fact that $W^{(h)}_{\rho,x}(\bar{t},\omega)$ is 
the Wigner function for hole excitations at time $\bar{t}$ and energy $\hbar\omega$. Consequently,
the upper bound on $W_{\rho,x}^{(e)}(\bar{t},\omega)$ is also required for a probabilistic
interpretation of the hole  Wigner distribution function.  For a probability distribution in the $(\bar{t},\omega)$ plane,
this upper bound expresses the Pauli principle.

\medskip

In full generality, the Wigner function for conduction electrons within a metal 
contains the Fermi sea: $W^{(e)}_{\rho,x}(\bar{t},\omega)\to 1$ for sufficiently negative $\omega$
and $W_{\rho,x}(\bar{t},\omega)\to 0$ at high enough energy. In between, the Wigner function can get various values, sometimes 
negative or strictly larger than one. In such cases, it cannot be interpreted as a probability distribution. We shall come back 
on this issue while discussing explicit examples in the next section.

\medskip

Note that a $T$-periodic system generates a single electron distribution invariant in $(t,t')\mapsto (t+T,t'+T)$
which can thus be decomposed as a Fourier transform
with respect to $\tau=t-t'$ and a Fourier series with respect to $\bar{t}=(t+t')/2$:
\begin{equation}
\label{eq:HBT:tomography:G-periodic}
\mathcal{G}_{\rho}^{(e)}(t,t')=\sum_{n=-\infty}^{+\infty}e^{-2\pi i nf\bar{t}}\int \mathcal{G}^{(e)}_{\rho,n}(\omega)
e^{-i\omega \tau}\frac{d\omega}{2\pi}\,
\end{equation}
where $f=1/T$ is the driving frequency. This single electron coherence leads to
a $T$-periodic Wigner function: $W^{(e)}_\rho(\bar{t}+T,\omega)=W_\rho^{(e)}(\bar{t},\omega)$
whose expression as a Fourier series reads:
\begin{equation}
\label{eq:Wigner:T-periodic}
W^{(e)}_\rho(\bar{t},\omega)=\sum_{n=-\infty}^{+\infty} v_F\mathcal{G}_{\rho,n}^{(e)}(\omega)\,e^{-2\pi inf\bar{t}}\,
\end{equation}
in which the position $x$ has been dropped out for simplicity.
As we shall see, this expression is of great use in numerical evaluations of the Wigner function in the framework of
Floquet scattering theory.

\medskip

Integrating the Wigner function of a quantum particle over position or momentum gives
the probability distribution of the conjugated variable. Here, partial integrals of the Wigner function
give access to physically relevant quantities such as the average excess current with
respect to a chemical potential $\mu$:
\begin{equation}
\label{eq:Wigner:average-current}
\langle i(x,\bar{t})\rangle_\rho = -e\int \Delta W_{\rho,x}^{(e)}(\bar{t},\omega)\,\frac{d\omega}{2\pi}\,
\end{equation}
where $\Delta W_{\rho,x}^{(e)}(\bar{t},\omega)$ denotes the excess Wigner function with respect to the
Wigner function $\Theta(\mu/\hbar-\omega)$ of the Fermi sea at chemical potential $\mu$.
Note that measuring this quantity requires broadband high frequency 
measurements\cite{Gabelli:2006-1,Bocquillon:2012-2}.
In the same way, averaging over time gives access to the 
electronic distribution function at position $x$:
\begin{equation}
\label{eq:Wigner:electron-distribution}
f_e(\omega,x)=\frac{1}{T}\int_{-T/2}^{T/2} W_{\rho,x}^{(e)}(\bar{t},\omega)\,d\bar{t} \,.
\end{equation}
This quantity can be measured using dc current measurements through an adjustable 
energy filter\cite{Altimiras:2010-1}. Note that the Wigner function for electrons in a 
conductor does not satisfy the normalization condition
satisfied by the Wigner function of a single quantum particle: integrating
the excess Wigner distribution $\Delta W^{(e)}_{\rho,x}(\bar{t},\omega)$ with 
respect to a given chemical potential  in the $(\bar{t},\omega)$ plane gives 
the total excess charge in $-e$ units. 

Finally, let us point out that
as of today, there is no way to directly access the value of the Wigner function 
at a given point $(t,\omega)$ in a quantum conductor. By contrast, such direct measurements 
of the Wigner function in cavity QED
are possible and have indeed been performed\cite{Bertet:2002-1} but they
rely on the measurement of the parity of the photon number\cite{Lutterbach:1997-1}, 
the equivalent of which 
is, as far as we know, not accessible in electron quantum optics. The problem of reconstructing
the Wigner function for electrons in quantum Hall edge channels through interferometry experiments 
will be discussed in section \ref{sec:interferometry}.

\section{Examples}
\label{sec:examples}

Let us now discuss several important examples of single electron coherences emitted by
various electronic sources. We shall first consider the single electron coherence emitted by
a driven Ohmic contact.

\subsection{Voltage drives}
\label{sec:Wigner:pulses}

\subsubsection{General properties}
\label{sec:Wigner:pulses:general}

The single electron coherence emitted by an ideal Ohmic contact driven by a time dependent voltage
$V(t)$ is:
\begin{equation}
\mathcal{G}_V^{(e)}\left(t+\frac{\tau}{2},t-\frac{\tau}{2}\right) = 
\exp{\left(\frac{ie}{\hbar}\int_{t-\frac{\tau}{2}}^{t+\frac{\tau}{2}}V(\tau')\,d\tau'\right)}\mathcal{G}_\mu^{(e)}(\tau)\,.
\end{equation}
In the case of a $T$-periodic potential, the single electron coherence can be
obtained in terms of the photo assisted transition amplitudes $p_l[V_{\mathrm{ac}}]$
associated with the a.c. component of the drive\cite{Dubois:2013-1}. 
The resulting Wigner function is then expressed as
\begin{eqnarray}
\label{eq:Wigner:voltage:Floquet}
W^{(e)}(t,\omega) & = & \sum_{(n_+,n_-)\in \mathbb{Z}^2}p_{n_+}[V_{\mathrm{ac}}]p_{n_-}[V_{\mathrm{ac}}]^*\,
e^{2\pi i(n_--n_+)ft}\nonumber \\
& \times & f_{\bar{\mu}}(\omega -\pi (n_++n_-)f)\,
\end{eqnarray}
where $\bar{\mu}=\mu-eV_{\mathrm{dc}}$ denotes the chemical potential 
shifted by the d.c. component of the drive. The electronic
occupation number is obtained by averaging Eq.~\eqref{eq:Wigner:voltage:Floquet} 
over the time $t$. This selects terms with $n_+=n_-$, thus leading to 
\begin{equation}
f_e(\omega)=\sum_{n\in \mathbb{Z}}|p_n[V_\mathrm{ac}]|^2f_{\bar\mu}(\omega-2\pi nf)\,.
\end{equation}
This expression clearly show the interpretation of $|p_n[V_{\mathrm{ac}}]|^2$ as the probability of
photo assisted transition. However, the Wigner function contains more information
than the photo assisted probabilities. These are the terms with 
$n_+\neq n_-$ in Eq.~\eqref{eq:Wigner:voltage:Floquet}
which are sensitive to the phases of the photo assisted transition amplitudes. They 
play a crucial role in ensuring that the current noise of the driven channel 
is equal to the current noise at thermal equilibrium
as is expected for a coherent states of the edge magnetoplasmon modes\cite{Grenier:2013-1}. 

\medskip

In the case of a sinusoidal drive $V_{\mathrm{ac}}(t)=V_0\,\cos{(2\pi ft)}$ at 
frequency $f$, $p_n[V_{\mathrm{ac}}]=J_n(eV_0/hf)$ where $J_n$ denotes 
the Bessel function of order $n$. The Wigner function is then obtained as:
\begin{equation}
\label{eq:Wigner:sinusoidal}
 W^{(e)}(t,\omega) = 
 \sum_{n \in \mathbb{Z}} \frac{ J_{n}\left(
 \frac{2eV_{0}}{hf} \cos{(2\pi ft)}\right)}{e^{\beta_{\mathrm{el}}\hbar(\omega+\pi nf)}+1}
\end{equation}
where $\beta_{\mathrm{el}}=1/k_BT_{\mathrm{el}}$. 
At zero temperatures, this Wigner function  exhibits singularities in the variable 
$\omega$ each time $\hbar\omega$ is a multiple of $hf/2$. 
Quantum effects are expected to be dominant in the regime of low temperature 
and low photon number. On the contrary, for
large photon number and high temperature, quantum features are expected to 
be small. Let us now turn to these two limiting regimes
of small $eV_0\ll hf$ and large amplitudes $eV_0\gg hf$.

\subsubsection{Small amplitudes}
\label{sec:Wigner:pulses:small}

The regime of small amplitudes is most suitably discussed in the sinusoidal case. 
Then, $hf$ represents the energy of photons absorbed or emitted by the electron 
gas and the condition $eV_0\ll hf$ expresses that the physics is
dominated by single photon processes.
In this regime, the first order contribution in $eV_0/hf$ to the Wigner function is:
\begin{equation}
\left.
\frac{\partial W^{(e)}(t,\omega)}{\partial (eV_0/hf)}
\right|_{V_0=0}=F_{\mu,T_{\mathrm{el}}}(\omega)
\cos{(2\pi ft)}\,
\end{equation}
where $F_{\mu,T_{\mathrm{el}}}(\omega)=
f_{\mu,T_{\mathrm{el}}}(\omega-\pi f)-f_{\mu,T_{\mathrm{el}}}(\omega+\pi f)$ 
is, at zero temperature, the characteristic function of the interval 
$[\hbar^{-1}\mu-\pi f,\hbar^{-1}\mu+\pi f]$.
Note that this contribution does not affect the electronic occupation number 
$f_e(\omega)$ but contributes to the average current by $(e^2/h)V(t)$ as expected 
from the linear response of a chiral edge channel. However, as we shall see
in section \ref{sec:interferometry:HBT-HOM}, it is instrumental to the recently 
proposed single electron tomography protocol\cite{Degio:2010-4}.

The first non trivial contribution to the electron distribution function arises at 
second order in $V_0$ and corresponds to processes in which a single photon is 
absorbed to promote one electron from the Fermi sea above the Fermi level:
\begin{equation}
\left.
\frac{\partial^2 W^{(e)}(t,\omega)}{\partial^2 (eV_0/hf)}
\right|_{V_0=0}=g_{\bar{\mu},T_{\mathrm{el}}}(\omega)\,\cos^2{(2\pi ft)}
\end{equation}
where $g_{\bar{\mu},T_{\mathrm{el}}}(\omega)=
f_{\bar\mu,T_{\mathrm{el}}}(\omega+2\pi f)+
f_{\bar\mu,T_{\mathrm{el}}}(\omega-2\pi f)-2f_{\bar\mu,T_{\mathrm{el}}}(\omega)$ is, at zero temperature, 
equal to $1$ for $\bar\mu< \hbar\omega\leq \bar\mu+hf$, to
to $-1$ when $\bar\mu-hf\leq \hbar\omega< \bar\mu$ and vanishes everywhere else.

At higher amplitudes, multiphotonic processes contribute. At zero temperature, 
the Wigner function exhibits singularities for $\omega$ multiple of $\pi f$ but only
the even multiples contribute to singularities in the occupation number as expected
from the theory of photon-assisted noise\cite{Lesovik:1994-1}.

\subsubsection{Large amplitudes}
\label{sec:Wigner:pulses:large}

Let us now turn to the opposite regime of large voltage amplitudes where the physics is 
dominated by multiphotonic processes.  In this case, let us make the discussion slightly 
more general by considering a smooth time dependent periodic voltage drive
that varies on a scale $\Delta V=\mathrm{max}(V(t))-\mathrm{min}(V(t))$ over a time 
scale $T=1/f$ where $f$ denotes the driving frequency.

\begin{figure}
\subfigure[\label{fig:Wigner:sinus:t0}\ $T_{\mathrm{el}}=0\ \mathrm{K}$]{
\input{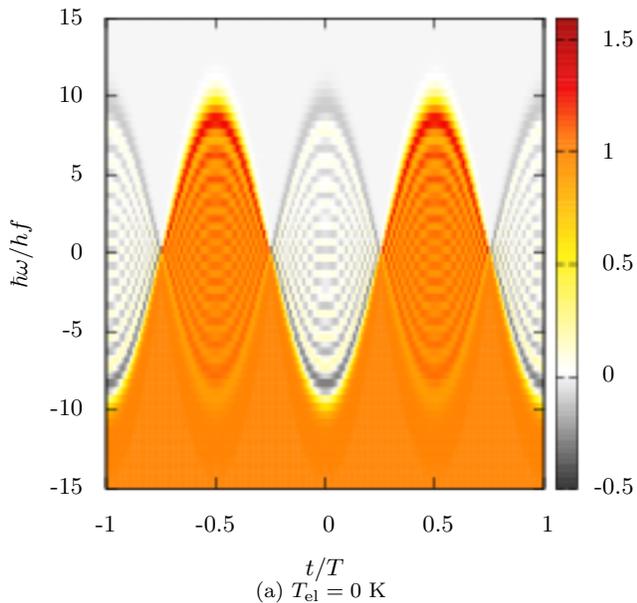}}
\subfigure[\label{fig:Wigner:sinus:tnz}\ $T_{\mathrm{el}}=0.25eV_{0}/k_{B}$]{
\input{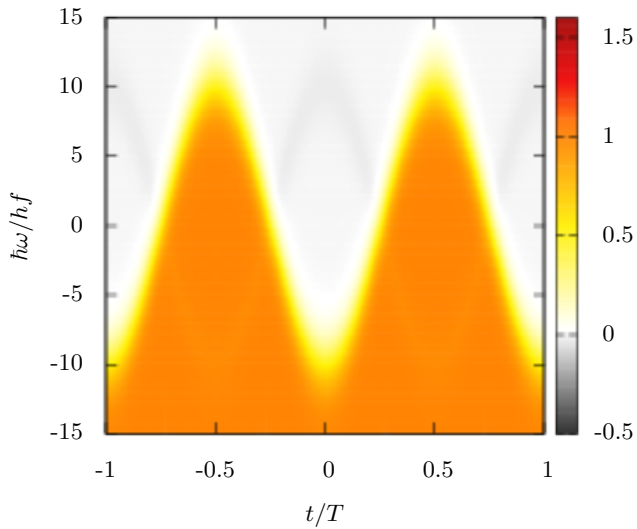}}
\caption{\label{fig:Wigner:sinus} Density plot of the Wigner function for
a sinusoidal drive $V(t)=V_{0}\cos{(2\pi ft)}$ in terms of $t/T$ and 
$\hbar\omega/hf$ for $eV_{0}/hf=20$ at (a) zero temperature and (b)
$k_{B}T_{\mathrm{el}}=eV_{0}/4$. Quantum ripples
discussed in the text are visible at $T_{\mathrm{el}}=0\ \mathrm{K}$.
The pixelization corresponding to the scale $\hbar\omega/hf=0.5$ is also only
visible at zero temperature.}
\end{figure}

To discuss the features of the Wigner function on energy scales of the order $e\Delta V$, we have to consider
$\mathcal{G}^{(e)}(t+\tau/2,t-\tau/2)$ over time scales such that $|f\tau|\ll 1$. We can then assume that $V(\tau')$ is
constant and equal to $V(t)$ between $t-\tau/2$ and $t+\tau/2$. This immediately leads to an adiabatic expression
for the Wigner function as a time-dependent Fermi distribution:
\begin{equation}
\label{eq:Wigner:adiabatic-voltage}
W^{(e)}(t,\omega)\simeq f_{\mu,T_{\mathrm{el}}}(\omega +eV(t)/\hbar)\,
\end{equation}
corresponding to a time dependent chemical potential $\mu(t)=\mu-eV(t)$. However, at zero temperature, quantum
interference effects lead to quantum corrections to this expression. They arise from the time dependence of
the voltage drive over the interval $[t-\tau/2,t+\tau/2]$. Considering a time $t$ such that $V''(t)\neq 0$, the
$\omega$ dependence of $W^{(e)}(t,\omega)$ exhibits  a Fresnel-like diffraction pattern. At
zero temperature, assuming that $V''(t)>0$, we find
\begin{equation}
\label{eq:Wigner:Vdrive:Airy-integral:positive}
W^{(e)}(t,\omega) \simeq \int_{\frac{\omega+eV(t)}{\delta\omega(t)}}^{+\infty}\mathrm{Ai}(x)\,dx
\end{equation}
where $\mathrm{Ai}(x)$ denotes the Airy function\cite{Book:GradRyz} and
\begin{equation}
\label{eq:Wigner:Vdrive:Airy-integral:delta-omega}
\delta\omega(t)=\frac{1}{2}\left(\frac{e|V''(t)|}{\hbar}\right)^{1/3}\,.
\end{equation}
The Wigner function thus exhibits quantum ripples on the Fermi plateau around the value one and an exponential decay 
at high energy. These ripples are clearly visible on Fig.~\ref{fig:Wigner:sinus:t0} presenting the Wigner function 
generated by a sinusoidal drive for $eV_{0}/hf=20$ at zero temperature. On this figure, they appear as oscillations on
the top part of the waves of the driven Fermi sea (the semi classical potential is $\mu(t)=\mu-eV(t)$ and therefore $V''(t)>0$ 
corresponds to $\mu''(t)<0$)... Due to them, the Wigner function can be greater than one. 
For $V''(t)<0$, a similar computation shows that
\begin{equation}
\label{eq:Wigner:Vdrive:Airy-integral:negative}
W^{(e)}(t,\omega) \simeq \int_{-\infty}^{-\frac{\omega+eV(t)}{\delta\omega(t)}}\mathrm{Ai}(x)\,dx\,.
\end{equation}
On Fig.~\ref{fig:Wigner:sinus:t0}, the corresponding ripples correspond to the oscillations in the bottom
of the wave part of the driven Fermi sea. They lead to negative values of the Wigner function.

The energy scale $\hbar\,\delta\omega(t)$ associated with these ripples compares to $hf$ through
\begin{equation}
\frac{\hbar\,\delta\omega(t)}{hf}=\frac{1}{4\pi}\,
\left(\frac{2\pi eT^2|V''(t)|}{hf}\right)^{1/3}\,.
\end{equation}
For a moderately varying voltage such as a sinusoidal drive,
$T^2V''(t)$ is of the order of the total drive amplitude $\Delta V$. 
Therefore in the case of a large amplitude $e\Delta V\gg hf$, the scale 
$\hbar\,\delta \omega(t)$ is significantly larger than $hf$ as can already be seen on 
Fig.~\ref{fig:Wigner:sinus:t0}. 

When $\hbar\,\delta\omega(t)\gg hf$, this energy scale gives the 
temperature above which the quantum ripples disappear. A convenient
way to understand finite temperature effects is to remember that 
thermal fluctuations will smooth the Wigner function
over an energy scale equal to $k_BT_{\mathrm{el}}$.  The 
quantum ripples are thus expected to dispappear 
at finite temperature, when $k_BT_{\mathrm{el}}\gtrsim \hbar\,\delta\omega(t)$. In this regime, 
the adiabatic result given by Eq.~\eqref{eq:Wigner:adiabatic-voltage} is recovered as
can be seen from Fig.~\ref{fig:Wigner:sinus:tnz}.

Finally, one might consider a voltage drive that is strongly 
peaked around specific times. In this case, the local
energy scale $\hbar\,\delta\omega(t)$ might become of the order of 
$eV(t)$ itself and the Wigner function is locally
dominated by these interferences effect. In such a situation, one would 
not be able to see the overall picture of
the Fermi step at chemical potential $\bar{\mu}(t)=\mu-eV(t)$. As we shall see in
section \ref{sec:Wigner:sources:levitons}, Lorentzian pulses realize such a situation.

\subsection{Single electron sources}
\label{sec:Wigner:sources}

\subsubsection{The mesoscopic capacitor}
\label{sec:Wigner:sources:capacitor}

An on demand single electron source can be realized using a 
mesoscopic capacitor operated in the
non linear regime. This source has been demonstrated in
2007 by G.~Fève {\it et al}\cite{Feve:2007-1}. Properly operated, it emits a single electron and a single
hole excitation per period at GHz repetition rate. One of its main 
advantages is that these excitations are energy resolved
and that their average energies and width can be tuned to some extent. 

\medskip

A description of this source can be achieved within a non interacting electron approximation 
using the framework of the Floquet scattering theory. 
This approach has been developed by Moskalets and Büttiker to describe 
quantum mechanical pumping in mesoscopic
conductors\cite{Moskalets:2002-1} and since then has been applied to a variety of 
systems among which the mesoscopic capacitor\cite{Moskalets:2008-1}.
In particular it has been used to predict the low and finite frequency noise emitted by a 
periodically driven mesoscopic conductor\cite{Moskalets:2007-1,Moskalets:2013-2}. These theoretical results
have been compared to experimental results on finite frequency noise of the source\cite{Parmentier:2012-1}.
The Floquet scattering amplitude for electrons propagating through a 
driven quantum conductor is:
\begin{equation}
\label{eq:Floquet:2}
\mathcal{S}_{\mathrm{Fl}}(t,t')=\exp{\left(\frac{ie}{\hbar}\int_{t'}^{t}V_{d}(\tau)\,d\tau\right)}\,\mathcal{S}_{0}(t-t')\,
\end{equation}
where $V_{d}(\tau)$ is the periodic a.c. driving voltage applied to the dot and 
$\mathcal{S}_{0}(t-t')$ is the scattering amplitude accross the undriven conductor, expressed in real time. 
Knowing the Floquet scattering amplitude \eqref{eq:Floquet:2} leads to the real time 
single electron coherence emitted by the driven mesoscopic conductor\cite{Degio:2010-4} in terms
of the Floquet scattering amplitudes $\mathcal{S}_n(\omega)$ defined as:
\begin{equation}
c_{\mathrm{out}}(\omega)=\sum_{n=-\infty}^{+\infty}
\mathcal{S}_n(\omega)\,c_{\mathrm{in}}(\omega+2\pi nf)\,.
\end{equation}
A Fourier transform
then leads to the general expression for the Wigner function emitted by
a source described within the framework of Floquet scattering theory:
\begin{eqnarray}
\label{eq:Floquet:Wigner}
W^{(e)}(t,\omega)& = & \sum_{n,k=-\infty}^{+\infty}
\mathcal{S}_k(\omega +\pi nf)\mathcal{S}_{n+k}(\omega -\pi nf)^*\nonumber \\
&   & \times \,f_\mu(\omega +2\pi f(k+n/2))\,e^{-2\pi inft}\,.
\end{eqnarray}
This expression can then be used to compute either analytically or numerically the Wigner function within Floquet scattering
theory. 

\medskip

Let us now discuss the numerical results for the mesoscopic capacitor driven by a square voltage:
$V(t)=V_d$ for $0< t\leq T/2$ and $V(t)=-V_d$ for $T/2<t\leq T/2$.
The results presented here have been obtained for realistic values of the parameters of the mesoscopic capacitor.
We consider  $hf/\Delta=0.06$, $k_BT_{\mathrm{el}}/\Delta=0.01$ and $eV_d=\Delta/2$ so that the voltage step
corresponds to the level spacing of the dot.
These results have been obtained by evaluating the single electron coherence numerically using a specific
form for the Floquet scattering theory already used to interpret the experimental data\cite{Gabelli:2006-1}:
\begin{equation}
\mathcal{S}_0(\omega)=\frac{\sqrt{1-D}-e^{2\pi i\hbar(\omega-\omega_0)/\Delta}}{1-\sqrt{1-D}
\,e^{2\pi i\hbar(\omega-\omega_0)/\Delta}}\,.
\end{equation}
Here $\Delta$ denotes the level spacing within the dot and $D$ the dot to lead transmission 
controlling the tunneling between the dot and the chiral edge channel. Another control is the position  $\hbar\omega_0$ 
of the energy levels of the dot which can be controlled by applying a d.c. voltage to its top gate. Note that electron/hole
symmetry is realized when $\hbar\omega_0$ is an integer multiple of $\Delta$.
Depending on $D$, various behaviors are expected.

\begin{figure}
\input{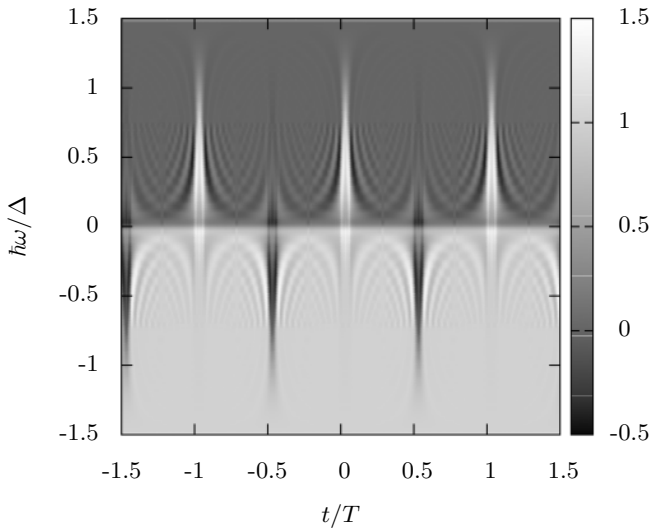}
\caption{\label{fig:ses:Wigner:D1} 
Density plot of the Wigner function emitted by the mesoscopic capacitor 
as a function of $t/T$ and $\hbar\omega/\Delta$ for 
$D=1$ for $hf/\Delta=0.06$, $k_BT_{\mathrm{el}}/\Delta=0.01$ and
a square voltage drive of amplitude $eV_d=\Delta/2$ in the symmetric situation 
($\omega_0=0$).}
\end{figure}

At $D=1$, electrons go around the dot only once and feel the effect of the voltage
drive during a very short time $\tau_0$ which is the time of flight around the dot. As shown on
Fig.~\ref{fig:ses:Wigner:D1}, excitation emission is concentrated at the times where
the voltage drive changes. This is expected since it is precisely when the drive changes that the
electrons going through the dot feel a sudden change of the electrical potential. 
Between two changes, the dot acts as a purely elastic scatterer and therefore we expect to see the emission of electrons
as if they were coming straight out of the reservoir. Consequently, the average current should be a succession 
of current pulses of duration $\tau_0$ corresponding to the sudden changes of the voltage drive.

From an edge magnetoplasmon perspective, the state generated by the 
mesoscopic capacitor at $D=1$ is a coherent state. It is therefore a coherent superposition 
of many electron/hole pairs and therefore we expect the single electron coherence to have an important contribution in 
the (e/h) quadrants. We then expect that excitations are created close to the Fermi level which is confirmed by 
Fig.~\ref{fig:ses:Wigner:D1}. 

\begin{figure}
\input{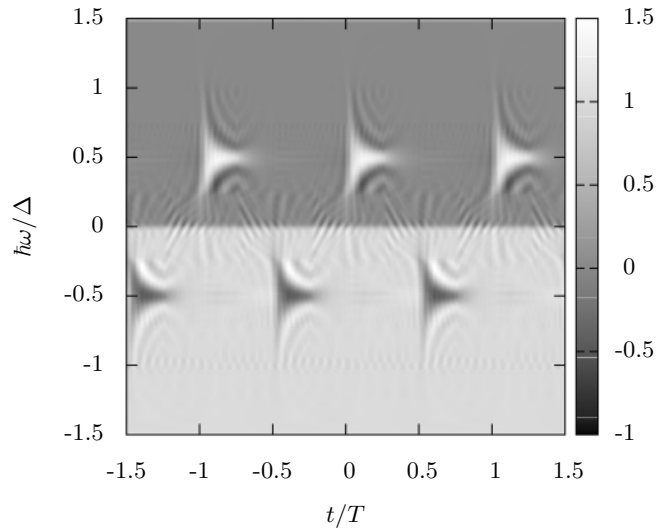}
\caption{\label{fig:ses:Wigner:optimal} 
Density plot of the Wigner distribution function emitted by the mesoscopic capacitor 
at $D=0.4$. for $hf/\Delta=0.06$, $k_BT_{\mathrm{el}}/\Delta=0.01$ and
a square voltage drive of amplitude $eV_d=\Delta/2$ in the symmetric situation 
($\omega_0=0$).}
\end{figure}

When $D$ decreases, the density of states within the dot 
becomes more and more textured\cite{Feve:2007-1}. Consequently we expect the source to emit 
electron and hole excitations that are better and
better defined and time shifted by a half period. 
Figure~\ref{fig:ses:Wigner:optimal} confirms this physical picture: it
clearly shows the succession of electronic and hole excitations emitted by the 
mesoscopic capacitor near its optimal point. 

The shape of these pulses can indeed be understood very simply by considering 
Lorentzian wave packets in energy, truncated to energies above the Fermi level:
\begin{equation}
\label{eq:lorentzian}
\widetilde{\varphi}_{e}(\omega)= 
\frac{\mathcal{N}_{\mathrm{e}}\,\Theta(\omega)}{\omega-\omega_{\mathrm{e}}-i\gamma_{\mathrm{e}}/2}
\end{equation}
where $\mathcal{N}_{\mathrm{e}}$ ensures normalization and $\gamma_{\mathrm{e}}$ denotes the electron 
escape rate from the quantum dot. For $|\hbar\omega_0|<\Delta$, we expect the electron to be emitted by
the mesoscopic capacitor at energy $\hbar\omega_{\mathrm{e}}=\Delta/2-\hbar\omega_0$ whereas the hole is
expected at energy $\hbar\omega_{\mathrm{h}}=\hbar\omega_0-\Delta/2$. The electronic escape rate is
then given by\cite{Mahe:2008-1,Nigg:2008-1} $\gamma_{\mathrm{e}}=2D\Delta/h(2-D)$.

\begin{figure*}
\input{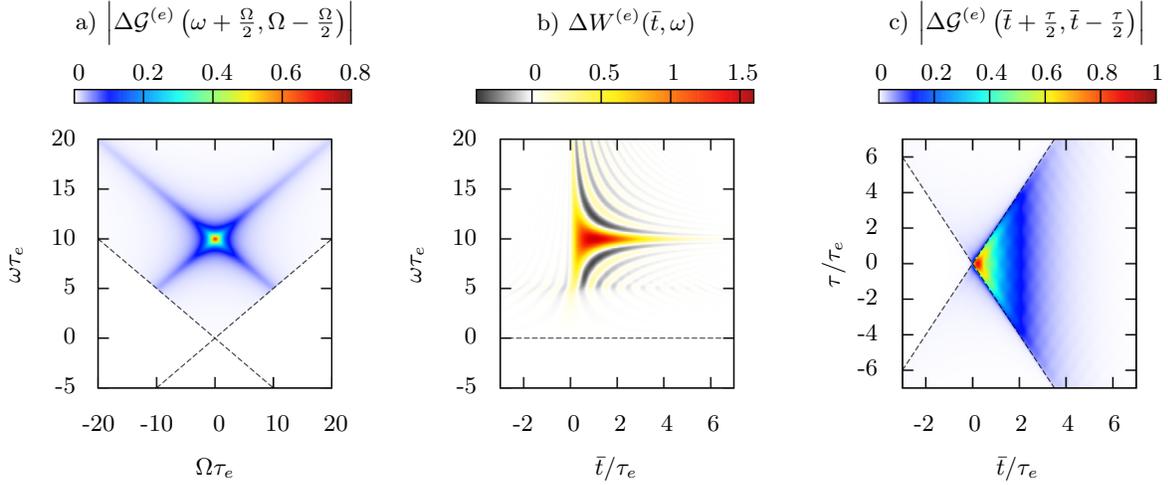}
\caption{\label{fig:triptique} Density plots associated with the three 
representations of single electron coherence: (a) the frequency domain representation 
(dashed lines delimitate the (e), (h) and (e/h) quadrants)
(b) the Wigner function representation (horizontal dashed line is the Fermi level)  and
(c) the frequency domain representation (dashed lines correspond to $t=0$ and $t'=0$).
These density 
plots correspond to a an energy resolved electronic wavepacket given 
by \eqref{eq:lorentzian} with $\omega_{\mathrm{e}}\tau_{\mathrm{e}}=10$.}
\end{figure*}

To understand the limit $\gamma_{\mathrm{e}}/\omega_{\mathrm{e}}\ll 1$, let us first neglect the 
truncation of the wave packet. Within this approximation, 
the associated excess Wigner function is given by:
\begin{equation}
\label{eq:wigner:wavepacket}
\Delta W^{(e)}(t,\omega)\sim 2\gamma_{\mathrm{e}} 
\Theta(t)\frac{\sin{(2(\omega-\omega_{\mathrm{e}})t)}}{\omega-\omega_{\mathrm{e}}}
\,e^{-\gamma_{\mathrm{e}}t}\,.
\end{equation}
This expression leads to a triangular shape in the 
$((\omega-\omega_{\mathrm{e}})/\gamma_{\mathrm{e}},\gamma_{\mathrm{e}}t)$ 
variables which reflects the Heisenberg time/energy
uncertainty principle: right after its emission, the spreading in energy is large 
and then becomes sharper. As expected the duration of the excitation is 
governed by $\tau_{\mathrm{e}}=1/\gamma_{\mathrm{e}}$. This Wigner function also exhibits
some negative values appearing as dark quantum ripples on Fig.~\ref{fig:triptique}(b) 
but apart from these, it is mainly a positive bump. 
Taking into account the truncation of the Lorentzian 
in Eq.~\eqref{eq:lorentzian} alters this image at low
energies: it leads to the vanishing 
of the single electron coherence 
outside the (e) quadrant as shown on Fig.~\ref{fig:triptique}(a). 
Consequently, for $\omega\lesssim \omega_{\mathrm{e}}/2$, the 
Fourier transform of $\Delta W^{(e)}(t,\omega)$ with respect
to $t$ is much smaller for $\omega \lesssim\omega_{e}/2$
as can be seen on Fig.~\ref{fig:triptique}(b). The residual interference pattern 
shows broader and fainter fringes
as $\omega$ goes to zero. These time oscillations at fixed $\omega$ arise 
from the residual coherence in the (e) quadrant ($\omega_{\pm}>0$) of the 
Fourier plane and $\omega_{+}+\omega_{-}=2\omega$.

This explains the band seen for $|\omega|\lesssim \Delta/2$
on Fig.~\ref{fig:ses:Wigner:optimal} for $|\omega|\lesssim |\omega_{\mathrm{e}}|/2$:  
the positive bump gives way to a fainter pattern of interferences fringes. 
This truncation effect, 
which can be interpreted as an expression of the Pauli principle, is of course 
sharper when $\gamma_{\mathrm{e}}\ll \omega_{\mathrm{e}}$. Its consequences in the 
time domain are discussed in Appendix~\ref{appendix:wavepacket}. The same remarks
apply to hole excitations which appear as dips in the Fermi sea.

\medskip

With $D$ going down closer to zero, the escape times of the electron and hole increase 
as one enters the shot noise regime of the source. When these
escape times become comparable to the half period $T/2$, the electron 
and the hole do not have the time to escape
before the voltage drive changes again. In this case, electron/hole coherences are 
expected to reappear 
since the capacitor generates a superposition of various electron/hole pair 
excitations of the form\cite{Degio:2010-4}
$|F\rangle +\psi^\dagger[\varphi_e]\psi[\varphi_h]|F\rangle$. 
The (e/h) coherences for such a superposition are of the form 
$\varphi_e(x)\varphi_{h}^*(y)$ and $\varphi_h(x)\varphi_{e}^*(y)$. 
In the Wigner distribution function, they appear at  
$\omega\simeq (\omega_{\mathrm{e}}+\omega_{\mathrm{h}})/2$ and oscillate in time
at angular velocity $\omega_{\mathrm{e}}-\omega_{\mathrm{h}}$ (remember that 
$\omega_{\mathrm{e}}>0$ for an electronic excitation and $\omega_{\mathrm{h}}<0$
for a hole excitation). We thus expect fast oscillations in time at 
mid position between the electron and
hole energies. 

\begin{figure}
\input{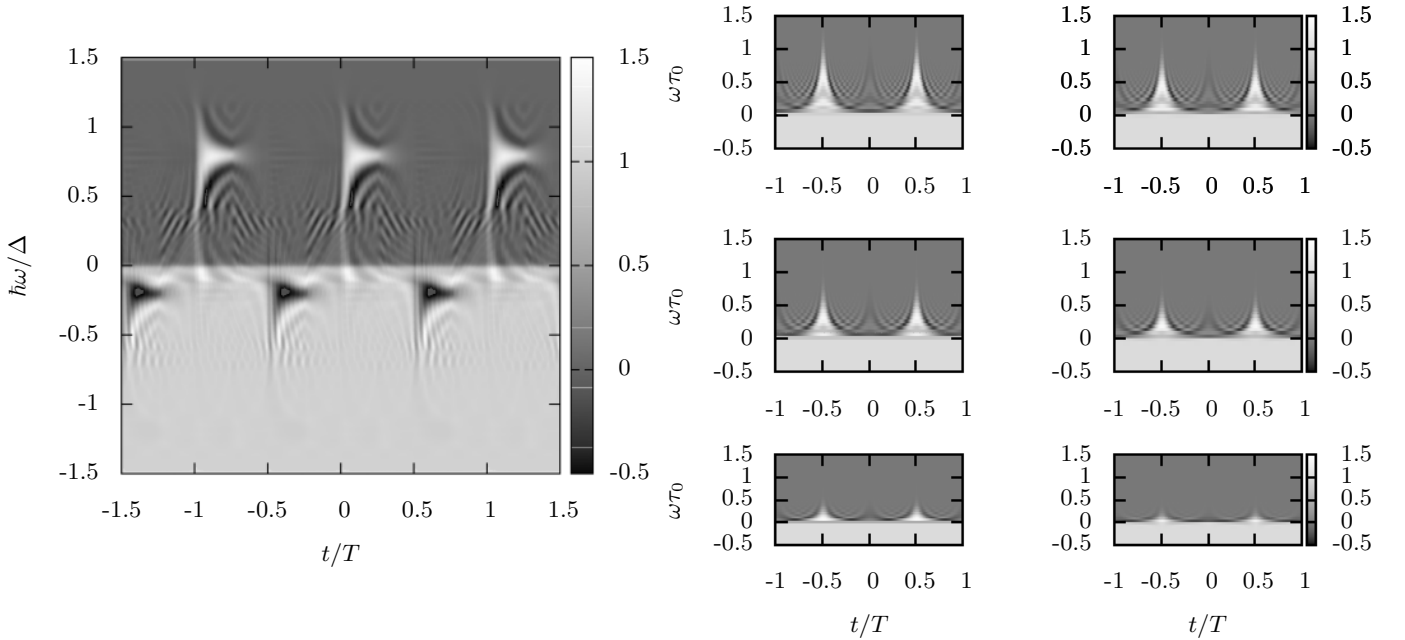}
\caption{\label{fig:ses:Wigner:shotnoise:shifted} 
Density plot of the Wigner function emitted by the mesoscopic capacitor 
as a function of $t/T$ and $\hbar\omega/\Delta$ for 
$D=0.4$, $k_BT_{\mathrm{el}}/\Delta=0.01$, $hf/\Delta=0.06$ and
when the energy levels of the dot at shifted by $0.3\Delta$.}
\end{figure}

To confirm this picture, Fig.~\ref{fig:ses:Wigner:shotnoise:shifted} depicts the Wigner 
function emitted by the mesoscopic capacitor in the low $D$ regime with a shift in the 
energy levels of the dot ($\omega_{0}\neq 0$). As expected, it  exhibits fast oscillations precisely
at the mid position between the electron and hole peaks. 
Note also the way the triangular shape of the 
hole excitation dips are truncated close to the Fermi surface due 
to the Pauli exclusion pinciple. 
Note that such oscillations were also
visible on Fig.~\ref{fig:ses:Wigner:optimal} but they are 
exactly at the Fermi energy since for this figure,
the capacitor is assumed to be operated at an electron/hole symmetric point. 
 
\subsubsection{Lorentzian pulses}
\label{sec:Wigner:sources:levitons}

More recently, another source of single to few electronic excitations in a 2DEG at zero
magnetic field has been developed\cite{Dubois:PhD}
based on ideas by Levitov, Ivanov, Lee and Lesovik\cite{Levitov:1996-1,Ivanov:1997-1}. 
The idea
is to use Lorentzian voltage pulses carrying a quantized charge in units of $e$ 
to generate purely electronic excitations. Experimentally, minimizing the partition
noise of these excitations is used to minimize the production of 
spurious electron/hole pairs\cite{Gabelli:2012-1,Dubois:2013-1}.
Contrary to the excitations studied in the previous paragraph, 
these electron pulses are time resolved instead of being energy resolved. Moreover, 
excitations carrying more
than an elementary charge under the form of a coherent wave packet 
of $n$ electrons can be generated.
More precisely, such an excitation is a Slater determinant built from the 
$1\leq k\leq n$ mutually orthogonal electronic wavefunctions given by\cite{Grenier:2013-1}:
\begin{equation}
\label{eq:coherence:Levitov:wavefunctions}
\varphi_{k}^{(\tau_0)}(\omega)=\sqrt{2\tau_0}\,\Theta(\omega)e^{-\omega\tau_{0}}L_{k-1}(2\omega\tau_0)\,
\end{equation}
where $\tau_0$ denotes the duration of the Lorentzian pulse and $L_k$ is the $k$th Laguerre 
polynomial\cite{Book:GradRyz}.
Their single electron coherence can be computed analytically 
in the case of a single pulse and also
from the Floquet scattering theory in the case of a periodic train of 
pulses\cite{Grenier:2013-1}. 

\medskip

For a single Levitov excitation of charge $-ne$ ($n\geq 1$) and 
width $\tau_{0}$, the associated excess Wigner
function is given by:
\begin{eqnarray}
\label{eq:Wigner:Levitov:n}
\Delta W^{(e)}(t, \omega)
& = & \sqrt{4\pi}\Theta(\omega) e^{-2 \omega \tau_{0}} \sum^{n-1}_{k=0} \sum^{k}_{l=0} 
\left(\frac{2\omega \tau_{0}}{\sqrt{\omega t} }\right)^{2l+1}
\nonumber\\
&   & \times \frac{L^{(2l)}_{k-l}(4\omega \tau_{0})}{l!}\,J_{l+\frac{1}{2}}(2\omega t)\,
\end{eqnarray}
where $J_{l+\frac{1}{2}}$ denotes the Bessel function of order $l+1/2$ and $L^{(m)}_{n}$ is 
the generalized Laguerre  polynomial\cite{Book:GradRyz} of order $n$. Note that for $n=1$, the wave 
function $\varphi_{0}^{(\tau_{0})}$ being exponential in energy, one
expects the Wigner distribution function to have a similar form as 
Eq.~\eqref{eq:wigner:wavepacket} up to the exchange of  $t$ and $\omega$.

However, in order to discuss the physics of these Lorentzian pulses, it is more convenient
to consider a periodic train of excitations since, 
in this case, non integer values of the charge can be  
considered\cite{Dubois:2013-1,Grenier:2013-1}. Using Eq.~\eqref{eq:Floquet:Wigner}, 
we have plotted the Wigner function of a train of Lorentzian pulses 
$f\tau_0=0.05$ at zero temperature and for increasing values of $\alpha$ which denotes
the average charge per pulse in units of $-e$. 

Figure \ref{fig:Levitov:Wigner},
show trains of time-resolved excitations that are not 
separated in energy from the Fermi level (compare with 
Fig.~\ref{fig:ses:Wigner:optimal}) as expected. Time-energy uncertainty 
leads to the spreading of the 
excitation in time close to the Fermi surface. 
Increasing the amplitude of the drive or 
equivalently $\alpha$, we see that these excitations grow in the energy direction. 
We also see the  appearance of maximas that are indeed the quantum ripples discussed
in section \ref{sec:Wigner:pulses:large}. But 
in the present case, they are more prominent since we are not
in an adiabatic limit. More interesting, we see that the Fermi 
sea is left pristine each time $\alpha$ is an integer confirming that
for positive integer $\alpha$, the source is emitting a strain of purely electronic excitations. 

On the contrary, when $\alpha>0$ is not an integer,
hole excitations are expected since in this case, each pulse should be understood as a collective excitation. Comparing 
$\alpha=1/2$, $3/2$ and $5/2$, we see that the hole contribution diminishes: this is not surprising in view of the Pauli
principle since one adds a half electronic excitation on top of a Slater determinant in which more and more
electronic states close to the Fermi level are populated. 
 
\begin{figure}
\input{Figures-tex/figure-levitons.tex}
\caption{\label{fig:Levitov:Wigner} 
Density plot of the Wigner functions of trains of Lorentzian pulses
of width $\tau_0$ such that $f\tau_0=0.05$, at zero electronic temperature and
for increasing values of $\alpha$: graphs on the left correspond to $\alpha=1$, $2$ and $3$ and
graphs on the right to $\alpha=0.5$, $1.5$ and $2.5$. The same colormap applies to all the graphs.}
\end{figure}

\section{Interferometry}
\label{sec:interferometry}

Let us now discuss interferometry experiments, starting first with single particle interferences 
(Mach-Zehnder interferometry) and then discussing two particle interferometry 
experiments based on the Hanbury Brown Twiss effect.

\subsection{Mach-Zehnder interferometry}
\label{sec:interferometry:MZI}

\begin{figure}
\includegraphics[width=8cm]{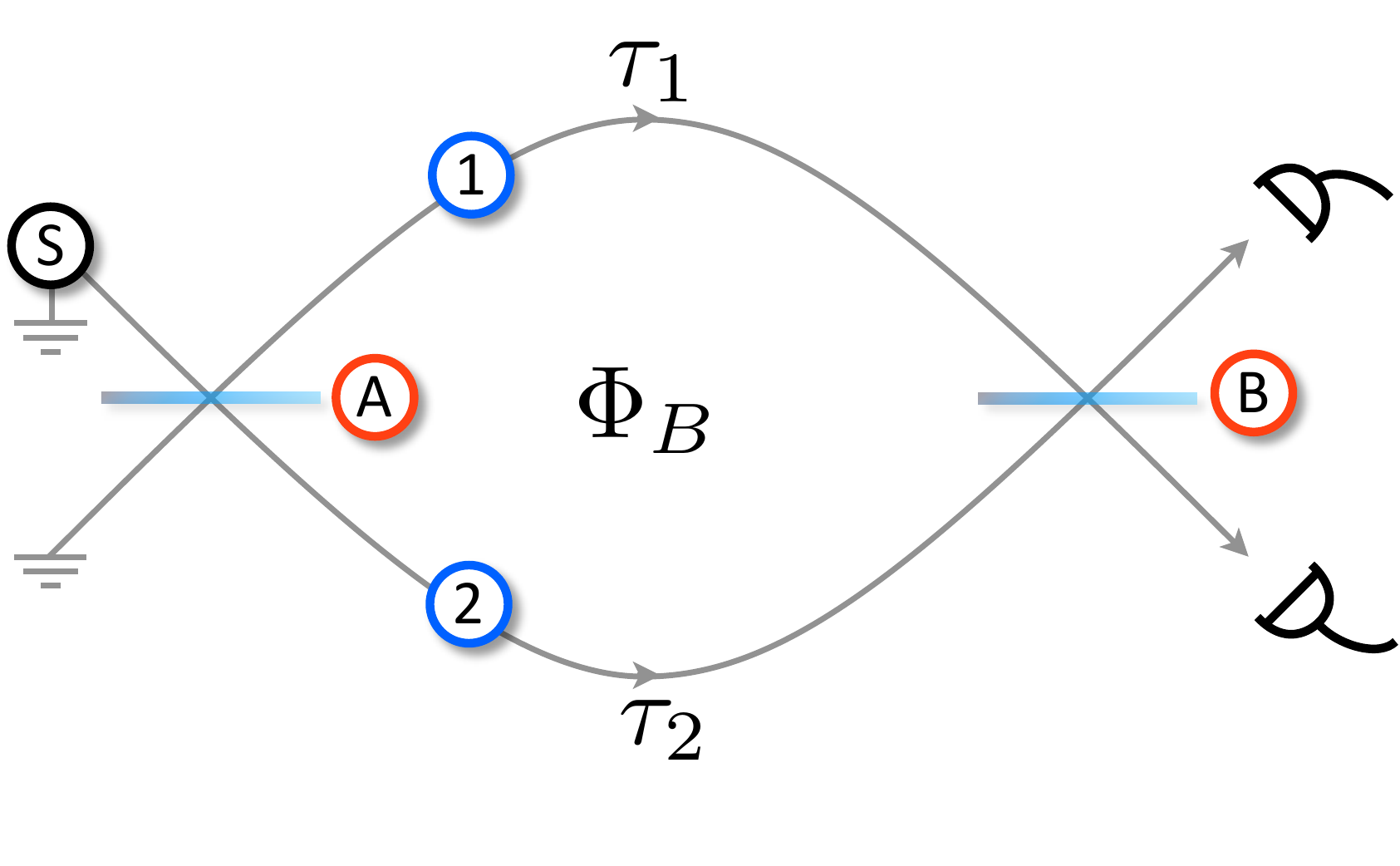}
\caption{\label{fig:MZI} Schematic view of a Mach-Zehnder interferometer: 
two incoming channels arrive on a beam splitter $A$ and fly along two 
paths 1 and 2 with respective times of flights $\tau_1$ and $\tau_2$ 
enclosing a magnetic flux $\Phi_B$ and recombine at a beam splitter $B$. An electronic
source (S) is places on the incoming channel 1.}
\end{figure}

We consider a Mach-Zehnder interferometer built from two QPCs $A$ and $B$
(see Fig.~\ref{fig:MZI}) whose scattering matrices $S^{(j)}$ ($j=A,\ B$) are of the form
\begin{equation}
S^{(j)}=\left(
\begin{array}{cc}
\sqrt{1-T_j} & i\sqrt{T_j}\\
i\sqrt{T_j} & \sqrt{1-T_j}
\end{array}
\right)\,.
\end{equation}
The two arms of the MZI encircle a region threaded by a magnetic flux 
$\Phi_B=\phi_B\times (h/e)$. The length of the two arms of the MZI are 
$l_{1,2}$ and here, we assume that electronic propagation is ballistic 
and non dispersive within each arm, thus leading to respective times 
of flights $\tau_{1,2}$. An electronic source (S) is located onto the 
incoming channel $1$ and both incoming 
channels are at chemical potential $\mu$ when the source is off. 
In this case, the outcoming channel are at the same chemical potential $\mu$. 

In full generality, a source generates an excess single electron coherence 
that propagates through the MZI. The excess Wigner function in the 
outcoming channel $1$ is then given by
\begin{eqnarray}
\label{eq:MZI:Wigner}
\Delta W^{(e)}_{1,\mathrm{out}}(t,\omega) & = & \sum_{j=1,2}\mathcal{M}_{j,j}\Delta W^{(e)}_{1,\mathrm{in}}(t-\tau_j) \\
& + & 2|\mathcal{M}_{1,2}|\cos{(\omega \tau_{12}+\phi)}\,\Delta W^{(e)}_{1,\mathrm{in}}(t-\bar{\tau},\omega)\,\nonumber
\end{eqnarray}
where $\tau_{12}=\tau_1-\tau_2$ and $\bar{\tau}=(\tau_1+\tau_2)/2$ respectively denote the
difference and the average of the two times of flights and $\phi=\mathrm{Arg}(\mathcal{M}_{12})+2\pi\phi_B$
accounts for the magnetic phase. Note that this is the Wigner function version of the 
discussion of the ideal MZI interferometer by Haack {\it et al}\cite{Haack:2012-2}.
The coefficients $\mathcal{M}_{i,j}$ are associated with the beam splitters
and, in the present case, are given by:
\begin{eqnarray}
\mathcal{M}_{11} & = & (1-T_A)(1-T_B),\\
\mathcal{M}_{22} & = & T_AT_B,\\
\mathcal{M}_{12} & = & \sqrt{T_A(1-T_A)}\,\sqrt{T_B(1-T_B)}\,.
\end{eqnarray}
In Eq.~\eqref{eq:MZI:Wigner}, the first two terms represent the contribution 
of electrons traveling classically along the two arms of the interferometer 
whereas the last term represents quantum interference effects 
associated with propagation along both arms. This quantum term 
presents oscillations in $\omega$ taking place over a scale $2\pi/|\tau_{12}|$. 
Changing the magnetic flux through  the MZI interferometer sweeps 
the phase of these quantum oscillations. These quantum fringes in the Wigner 
function are indeed characteristic of quantum superpositions. They 
have been recently observed for Schrödinger cat states
in cavity QED\cite{Deleglise:2008-1} and they are now routinely 
seen in circuit QED  
experiments\cite{Wang:2009-1,Hofheinz:2009-1,Shalibo:2013-1}. 

\medskip

\begin{figure}
\subfigure[\label{fig:MZI:Wigner:SES:d2}\ Partly overlaping wavepackets: $\tau_{12}=2\,\tau_{\mathrm{e}}$]{
\input{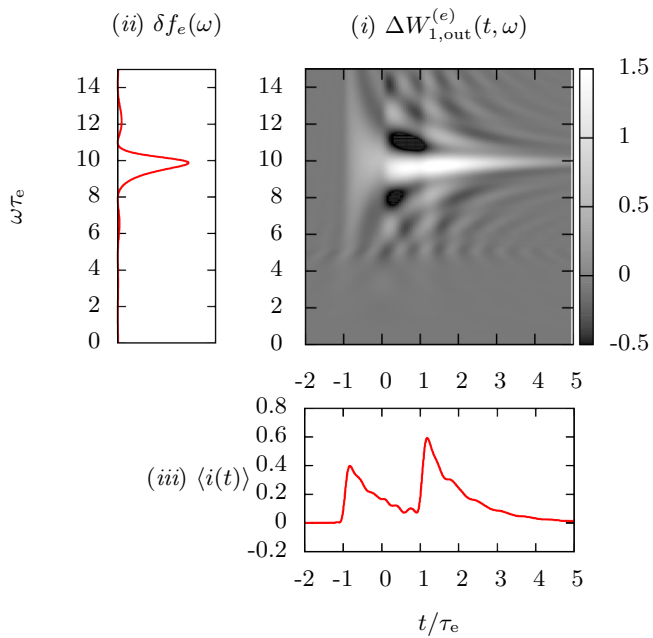}}
\subfigure[\label{fig:MZI:Wigner:SES:d9}\ Non-overlaping wavepackets: $\tau_{12}=9\,\tau_{\mathrm{e}}$]{
\input{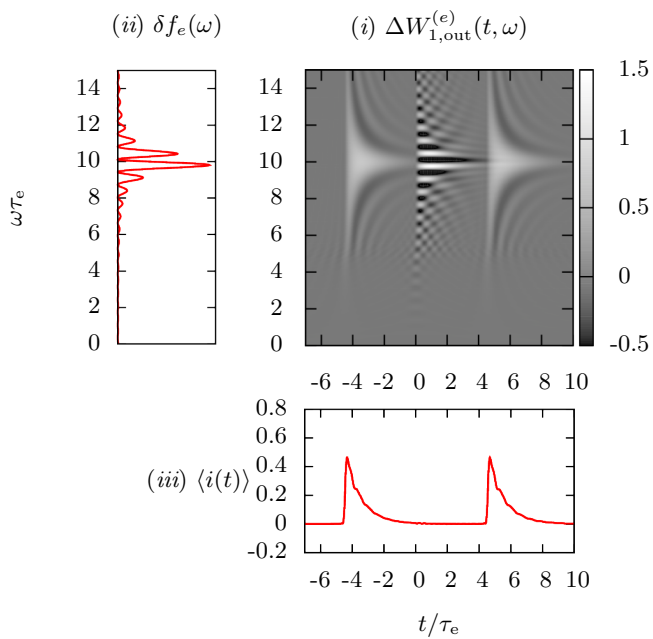}}
\caption{\label{fig:MZI:Wigner:SES} Excess single electron coherence at the output of an ideal Mach-Zehnder interferometer
for energy resolved wave-packets \eqref{eq:lorentzian} emitted at energy $\hbar\omega_{\mathrm{e}}$ and
with duration $\tau_{\mathrm{e}}=\gamma_{\mathrm{e}}^{-1}$ such that $\omega_{\mathrm{e}}\tau_{\mathrm{e}}=10$ and two differences of
time of flights $\tau_{12}$: ({\it i}) Density plot of the excess Wigner function 
$\Delta W^{(e)}_{1,\mathrm{out}}(t,\omega)$ given by Eq. \eqref{eq:MZI:Wigner} for $\phi=0$ 
as a function of $t/\tau_{\mathrm{e}}$ and $\omega\tau_{\mathrm{e}}$ 
({\it ii}) Excess electron distribution function $\delta f_e(\omega)$ in the outcoming channel (arbitrary units) ({\it iii})
average current $\langle i(t)\rangle$ in units of $-e/\tau_{\mathrm{e}}$.}
\end{figure}

Figure \ref{fig:MZI:Wigner:SES:d2} and \ref{fig:MZI:Wigner:SES:d9}
present the Wigner functions obtained by sending through an ideal MZI an energy resolved 
single electron wavepacket given by Eq.~\eqref{eq:lorentzian}. The two time shifted wave packets are
clearly visible. We have chosen the time of flight difference $\tau_{12}$ larger
than the duration of each wave packet so that the quantum fringes are clearly visible. 
In this regime, the two classical components of $\Delta W_{1,\mathrm{out}}^{(e)}(t,\omega)$
are clearly separated from the quantum interference features. Due to the quantum interference
contribution, the Wigner function exhibits pronounced regions with negative values
thus preventing it from any quasi-classical interpretation. 

Although they could be observed in a full quantum tomography of the 
single electron coherence\cite{Degio:2010-4},
these fringes could also be observed through their impact on the marginal distributions 
of the Wigner function, that is
to say the average current (see Eq.~\eqref{eq:Wigner:average-current}) and the electron 
distribution function 
(see Eq.~\eqref{eq:Wigner:electron-distribution}). 

\medskip

In order to observe the effect on the average current, 
the typical scale $2\pi /|\tau_{12}|$ of quantum oscillations in the Wigner function 
must be comparable
or larger than the energy spread of the excess coherence of the source so that these
oscillations are not averaged to zero when integrating over $\omega$. This 
condition expresses that
quantum interferences can be observed in the time domain only when the difference of 
times of flights
$\tau_{12}$ is comparable or smaller than the coherence time of the source\cite{Haack:2011-1}.  
This is still slightly the case
on Fig.~\ref{fig:MZI:Wigner:SES:d9} since we have considered $\tau_{12}=2\,\tau_{\mathrm{e}}$. 
The average current then shows two overlapping peaks whereas the electron distribution
exhibit mainly one peak with some light structure. Changing the magnetic flux by half a quantum decreases
the size of the second peak of the average current and also decreases the size of the 
central peak of the electron distribution
function. 

On the other hand, in the case of a strongly unbalanced MZI, such as the one depicted on 
Fig.~\ref{fig:MZI:Wigner:SES:d9},
interferences cannot be observed using a time resolved measurement. The average current depicted
on Fig.~\ref{fig:MZI:Wigner:SES:d9}-(iii)
exhibits two peaks that correspond to the two classical contributions to the excess Wigner function.
These two peaks do not change when the magnetic flow is varied. In this case, a frequency resolved
detector able to restore the overlap between electronic wavepackets 
is more appropriate to reveal the interferences. As shown on Fig.~\ref{fig:MZI:Wigner:SES:d9}
the quantum fringes of the Wigner function can be revealed by measuring the electron distribution
function. This is the single electron version of the 
channeled spectrum observed in optical interferometers using white light. 

\medskip

In practice, observing these effects might be quite challenging 
due to decoherence within the interferometer
itself\cite{Roulleau:2007-2,Roulleau:2008-1,Roulleau:2008-2,Neder:2007-4} although a proper design
of the sample leads to partial protection against 
decoherence\cite{Altimiras:2010-2,Huynh:2012-1}. 
Moreover, observing a channeled spectrum 
requires an unbalanced interferometer in which $|\tau_{12}|$ is 
greater than the inverse of the typical 
energy spreading of the electrons sent into the interferometer. 
Therefore, the best way to observe a
channeled spectrum would be to use a non equilibrium distribution 
 spreading over a broad energy range such as a double step generated by
a QPC\cite{Altimiras:2010-1} or the one generated by an ac sinusoidal voltage.
A typical energy range of 100 to 500 $\mu\mathrm{eV}$ corresponds to 
frequencies from 24 to 120~GHz 
so that an imbalance of a few $\mu\mathrm{m}$ would be sufficient to 
observe a channeled spectrum. 
However, such an experiment would rely on a continuous stream of 
undistinguishable electrons.
In order to observe a channeled spectrum with a single electron source, 
a larger imbalance is required and
excitations that are spread in energy such as Lorentzian pulses\cite{Dubois:PhD} 
with short duration (typically 10~ps) may
be appropriate. An alternative approach is to use energy resolved 
excitations\cite{Feve:2007-1} and change their
energies to probe the interferences at the single electron level. But discussing the 
observability of these effects in a realistic situation would require computing the Wigner
distribution function at the output of a Mach-Zehnder interferometer in the presence
of interactions, which would go beyond the scope of the present paper.

\subsection{HOM and HBT interferometry}
\label{sec:interferometry:HBT-HOM}

\subsubsection{Noise and correlations from coherences}

\begin{figure}
\begin{center}
\includegraphics[width=8cm]{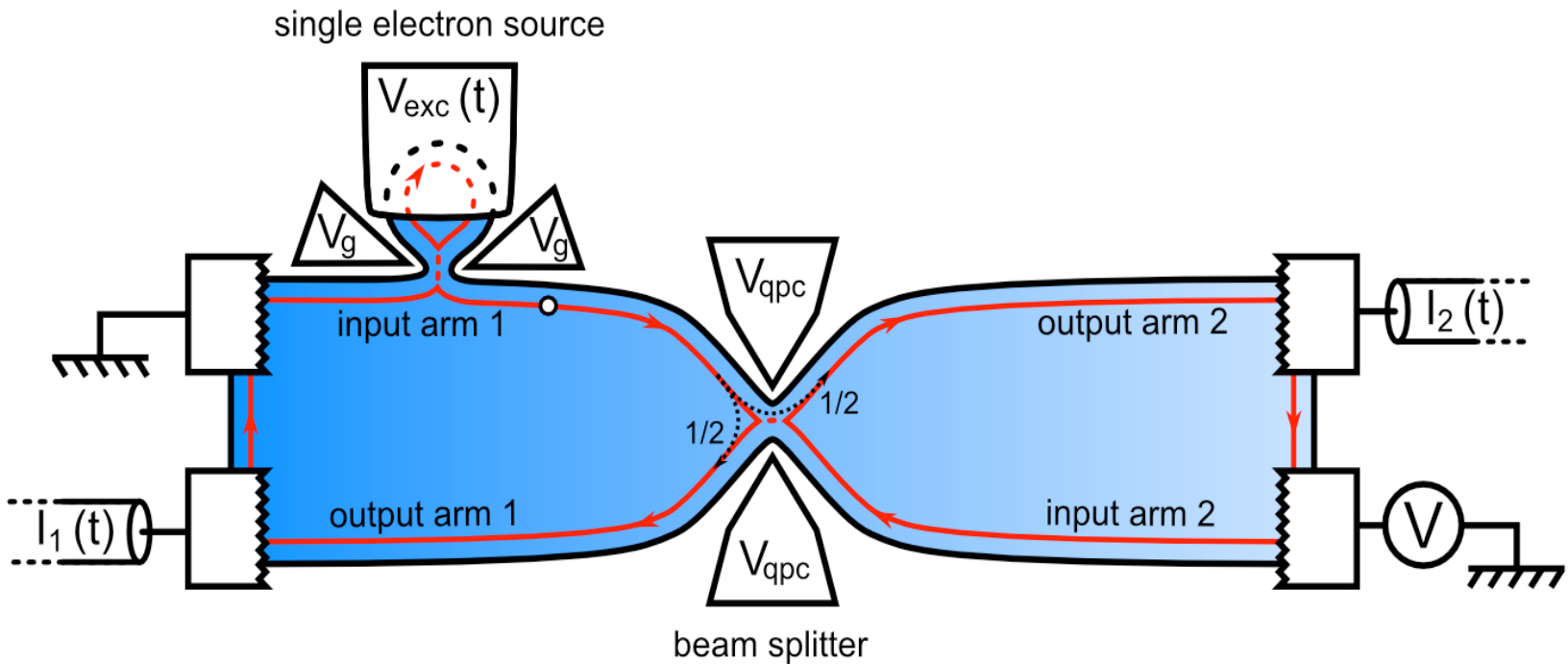}
\end{center}
\caption{\label{fig:HBT} Hanbury Brown \& Twiss setup for single electron tomography:
This setup is designed to characterize the single electron coherence of an on demand 
single electron source present on the incoming branch 1
close to the QPC (here $\mu_{1}=0$) and driven by the voltage $V_{\mathrm{exc}}(t)$. 
A reservoir with a time dependent chemical 
potential $\mu_{2}(t)=-eV(t)=\mu_{2}-eV_{\mathrm{ac}}(t)$ is connected to the incoming 
branch 2. One measures the low frequency correlation 
$S_{12}^{\mathrm{exp}}$ of the outcoming current $I_{1}$ and $I_{2}$.}
\end{figure}

\begin{figure}
\begin{center}
\includegraphics[width=8cm]{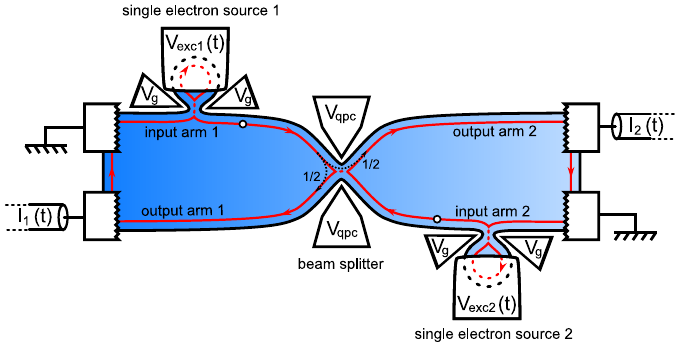}
\end{center}
\caption{\label{fig:HOM} Sketch of the usual Hong Ou Mandel experiment\cite{Bocquillon:2013-1}:
On demand single electron sources are present on each incoming branch 
and are driven by time shifted voltages: $V_{\mathrm{exc,2}}(t)=V_{\mathrm{exc,1}}(t-\Delta t)$. 
One measures the low frequency correlation  $S_{12}^{\mathrm{exp}}$ of the outcoming current $I_{1}$ and $I_{2}$.}
\end{figure}

Two particle interferences are at the heart of the Hanbury Brown Twiss (HBT) and Hong Ou Mandel (HOM)
experiments recently demonstrated with single electron excitations emitted by on demand single
electron source\cite{Bocquillon:2012-1,Bocquillon:2013-1}. They enable
us to imprint information about the single electron coherences into the current noise and correlations
issued from an electronic beam splitter of respective reflection and transmission probabilities $R$ and $T$
(see Figs.~\ref{fig:HBT} and \ref{fig:HOM}). This remark underlies the recently 
proposed single electron tomography protocol aimed at
reconstructing an unknown single electron coherence from noise measurements\cite{Degio:2010-4}. 
Predictions have been made for HOM noise signals within the Floquet scattering 
theory\cite{Olkhovskaya:2008-1,Moskalets:2013-1}, due to
electron and hole coherence\cite{Jonckheere:2012-1} 
and also for the $\nu=2$ edge channel system with 
short range interactions\cite{Wahl:2013-1}.

The aim of this paragraph is to revisit all these HBT interferometry experiments in terms of the Wigner 
function. We show that the Wigner function provides a simple and unified view of all these 
experiments and that it is the quantity of interest for their interpretation. 

\medskip

In the HBT interferometry setup depicted on Figs.~\ref{fig:HBT}
and \ref{fig:HOM}, the outcoming current correlations between edge
channels $\alpha$ and $\beta$ is defined as
\begin{equation}
\label{eq:HBT:1}
S_{\alpha\beta}^{\mathrm{out}}(t,t')=\langle i^{\mathrm{out}}_{\alpha}(t)\,
i^{\mathrm{out}}_{\beta}(t')\rangle -
\langle i^{\mathrm{out}}_{\alpha}(t)\rangle\langle i^{\mathrm{out}}_\beta(t')\rangle\,.
\end{equation}
It can be expressed in terms 
of the incoming ones $S_{\alpha\beta}^{\mathrm{in}}(t,t')$ and of a contribution $\mathcal{Q}(t,t')$
 coming from two particle interferences. 
This contribution $\mathcal{Q}(t,t')$ involves the incoming single particule coherences right
upstream the QPC:
\begin{equation}
\label{eq:HBT:Q}
\mathcal{Q}(t,t')=(ev_F)^2\left(
\mathcal{G}_{1}^{(e)}\mathcal{G}_{2}^{(h)}+
\mathcal{G}_{2}^{(e)}\mathcal{G}_{1}^{(h)}\right)(t',t)\,
\end{equation}
where, for HBT interferometry, $\mathcal{G}_{2}^{(e)}$ and  $\mathcal{G}_{2}^{(e)}$ 
are the coherences
emitted by a reservoir at chemical potential $\mu_2$ and electronic temperature $T_{\mathrm{el}}$.
The final expressions for outcoming current correlations are\cite{Degio:2010-4}:
\begin{eqnarray}
\label{eq:HBT:S11}
S_{11}^{\mathrm{out}} & = & R^2S_{11}^{\mathrm{in}}+T^2S_{22}^{\mathrm{in}} +RT\,\mathcal{Q}\\
\label{eq:HBT:S22}
S_{22}^{\mathrm{out}} & = & T^2S_{11}^{\mathrm{in}}+R^2S_{22}^{\mathrm{in}} +RT\,\mathcal{Q}\\
\label{eq:HBT:S12}
S_{12}^{\mathrm{out}} &  = & S_{21}^{\mathrm{out}} = RT\,(S_{11}^{\mathrm{in}}
 + S_{22}^{\mathrm{in}}-\mathcal{Q})\,
\end{eqnarray} 
where for simplicity we have omitted the $(t,t')$ arguments.

As discussed in Sec.~\ref{sec:eqo:G1:basics}, electron quantum optics differs from 
photon quantum optics on the nature of the vacuum. Even
when sources are switched off, a non vanishing single electron coherence is present and, 
at non zero temperature, leads to a non zero current
noise. We shall thus consider situations in which no source is switched on 
($\mathrm{off}/\mathrm{off}$), one of the two is switched on 
(cases $\mathrm{on}/\mathrm{off}$ and $\mathrm{off}/\mathrm{on}$) and both are 
switched  on ($\mathrm{on}/\mathrm{on}$). 
Denoting by $\Delta X$ the difference between $X$ in the situation under consideration 
and $X$ when the two sources are off, we obtain
\begin{subequations}
\begin{eqnarray}
\label{eq:HBT:on-off}
\Delta S_{11}^{\mathrm{on}/\mathrm{off}} &= & R^2\Delta S_{11}^{\mathrm{in}}
+RT(\Delta\mathcal{Q})_{\mathrm{on}/\mathrm{off}}\\
\Delta S_{11}^{\mathrm{off}/\mathrm{on}} &= & T^2\Delta S_{22}^{\mathrm{in}}
+RT(\Delta\mathcal{Q})_{\mathrm{off}/\mathrm{on}}
\label{eq:HBT:off-on}
\end{eqnarray}
\end{subequations}
where the HBT excess contribution in Eq.~\eqref{eq:HBT:on-off} is given by:
\begin{equation}
\label{eq:HBT:excess:1}
(\Delta\mathcal{Q})_{\mathrm{on}/\mathrm{off}}(t,t')= (ev_F)^2\left(\Delta\mathcal{G}_1^{(e)}
\mathcal{G}_2^{(h)}+\Delta\mathcal{G}_1^{(h)}\mathcal{G}_2^{(e)}\right)(t',t)
\end{equation}
with a similar expression for $(\Delta\mathcal{Q})_{\mathrm{off}/\mathrm{on}}(t,t')$. 
These expressions are relevant when one of the incoming channel
is populated by an equilibrium distribution. Since an equilibrium state is the analogous 
of the vacuum for photons in the electron quantum
optics channel, such a situation is analogous to the historical 
tabletop HBT experiment\cite{Hanbury:1956-2}. 
It has been recently
demonstrated with an on demand single electron source\cite{Bocquillon:2012-1}.

In the case of an HOM experiment (see Fig.~\ref{fig:HOM}), the two sources are switched on. In this case,
the excess outcoming noise $\Delta S_{11}^{\mathrm{on}/\mathrm{on}}$ is the sum of the
excess HBT noises and of an HOM contribution corresponding to two particle interferences between excitations
emitted by both sources: 
\begin{equation}
\label{eq:HBT:on-on}
\Delta S_{11}^{\mathrm{on}/\mathrm{on}}=\Delta S_{11}^{\mathrm{on}/\mathrm{off}}
+\Delta S_{11}^{\mathrm{off}/\mathrm{on}}+RT(\Delta\mathcal{Q})_{\mathrm{HOM}}\,.
\end{equation}
This HOM contribution (with respect to the situation where the two sources are switched off) is defined as:
\begin{equation}
\label{eq:HBT:excess:HOM}
(\Delta\mathcal{Q})_{\mathrm{HOM}}(t,t')=(ev_F)^2\left(\Delta\mathcal{G}_1^{(e)}\Delta\mathcal{G}^{(h)}_2+(1\leftrightarrow 2)\right)(t',t)\,.
\end{equation}
In electronic transport, we can only access to the time averaged component of current correlations at a given 
frequency $\Omega$:
\begin{equation}
\label{eq:HBT:exp}
S_{\alpha\beta}^{\mathrm{exp}}(\Omega)=
\int \overline{S_{\alpha\beta}^{\mathrm{out}}(\bar{t}+\tau/2,\bar{t}-\tau/2)}^{\bar{t}}
e^{i\Omega\tau}d\tau\,.
\end{equation}
Although the experiments are often performed at low frequency ($\Omega\simeq 0$), 
recent progresses in finite 
frequency noise measurements\cite{Parmentier:2010-1} motivate considering 
the case of finite frequency noise.
As a consequence, the HBT and HOM excess contributions to the experimental signals 
are given by overlaps in time of two excess coherences. Thanks to Plancherel's theorem, 
they can also be expressed as overlaps
of the corresponding Wigner functions. 
As we shall see, this makes the physical interpretation of the HBT and 
HOM two particle interferometry experiments extremely transparent.

\subsubsection{A Wigner view on the HBT and HOM experiments}

To begin with, let us first consider the HBT excess contribution given by 
Eq.~\eqref{eq:HBT:excess:1}.
Rewriting excess hole coherences in terms of 
electronic coherences and performing the time averaging and Fourier 
transform as in Eq.~\eqref{eq:HBT:exp} leads to
the excess HBT contribution of the outcoming noise under the form:
\begin{align}
\label{eq:HBT:overlaps}
(\Delta\mathcal{Q})_{\mathrm{on/off}}(\Omega)  = -
 e\langle i_1\rangle 
-  e^2\int \delta f_1(\omega)\,g_2(\omega,\Omega)
\frac{d\omega}{2\pi}
\end{align}
where $\langle i_1\rangle$ denotes the average dc current of source 1, 
$\delta f_1(\omega)=\overline{\Delta W_1^{(e)}(t,\omega)}^t$ is the 
excess electron distribution
function in the incoming channel 1 and
\begin{equation}
g_2(\omega,\Omega)=f_2(\omega-\Omega)+f_2(\omega+\Omega)\,
\end{equation}
is a double step going from zero for $\omega\gg |\Omega|$ to $2$ for 
$\omega\ll -|\Omega|$ through an intermediate
plateau at $1$ for $|\omega|\lesssim |\Omega|$. The steps are thermally broadened 
over an energy scale $k_{B}T_{\mathrm{el}}$.
Let us first remark that the result depends on the excess electron distribution function
$\delta f_1(\omega)$ which is precisely the time average part of single electron 
coherence in the incoming channel 1. 
This is  expected since we are measuring a time averaged quantity and the incoming 
channel 2 is populated with a stationary (equilibrium) state. 

The first contribution in the r.h.s. of Eq.~\eqref{eq:HBT:overlaps} corresponds to classical (Poissonian) 
partitioning of electrons and hole excitations sent in the incoming channel 1. This noise has
a white spectrum the corresponding term is independent from $\Omega$.

The second term in the r.h.s. of Eq.~\eqref{eq:HBT:overlaps} is an even function
of $\Omega$.
Its minus sign arises from the fermionic statistics of electrons and expresses antibunching. 
In fact, Eqs~\eqref{eq:HBT:Q} to \eqref{eq:HBT:S12} are also valid 
for bosons but the relation between
particle and anti-particle coherences would lead to a plus sign in the case 
of bosons in Eq.~\eqref{eq:HBT:overlaps}. In an experience with bosons, $g_{2}(\omega,\Omega)$ 
would arise the bosonic occupations numbers in a stationary reference beam arriving from input
arm 2. To understand the frequency dependence, we rewrite the average dc current 
in terms of the excess electron
distribution function. Eq.~\eqref{eq:HBT:overlaps} then becomes:
\begin{equation}
\label{eq:HBT:final-HBT}
(\Delta\mathcal{Q})_{\mathrm{on/off}}(\Omega)  =
e^2\int \delta f_1(\omega)\, (1-g_2(\omega,\Omega))\,\frac{d\omega}{2\pi}\,.
\end{equation}
This shows that at low zero temperature, the excess HBT contribution is given by 
the sum of electronic excitations
and of hole excitations whose energies exceed $\hbar |\Omega|$. This expression 
generalizes the formula
known for the outcoming current noise at zero frequency\cite{Bocquillon:2012-1}.

\medskip

Let us now discuss the case of an HOM experiment. The same line of reasoning shows that
the HOM contribution given by Eq.~\eqref{eq:HBT:excess:HOM} is the overlap of the
excess Wigner functions arriving from the two incoming channels:
\begin{align}
\label{eq:HBT:excess:HOM:exp}
(\Delta \mathcal{Q} )_{\mathrm{HOM}}(\Omega) = 
-  e^2\int \left[ \overline{\Delta W_1^{(e)}(t,\omega)\,\Delta W_2^{(e)}(t,\omega+\Omega)}\right.\nonumber \\
+  \left. \overline{\Delta W_2^{(e)}(t,\omega)\,\Delta W_1^{(e)}(t,\omega+\Omega)}^t\right]\frac{d\omega}{2\pi}\,.
\end{align}
This simple expression immediately suggests what happens physically: this 
term will contribute only when the excitations emitted by the two sources
overlap up to the frequency shift at which outcoming current noise or correlations are measured. 
Consequently, the noise and correlation in the HOM experiment contain an interesting
information about the single electron coherences injected in the HBT interferometer. 
Whereas the usual HOM experiment
gives access to the overlap of the excess Wigner functions of the two sources modulo an
experimentally controlled time shift, finite frequency measurements enable us to probe
a shift in energy. As a consequence, such measurements, although notably 
difficult\cite{Parmentier:2010-1}, could
be used to obtain qualitative information on relaxation mechanisms.

\medskip 

Finally, in the case of two ideal single electron sources, each one emitting a single electron wave 
packet per period characterized by its wave function $\varphi_{1,2}$,
the HOM contribution at low frequency reduces to\cite{Degio25}:
\begin{equation}
\label{eq:HBT:excess:HOM:wavepackets}
(\Delta \mathcal{Q})_{\mathrm{on}/\mathrm{on}}(\Omega=0)=-2e^2f\,|\langle \varphi_1|\varphi_2\rangle|^2\,.
\end{equation}
However, note that this interpretation of the HOM contribution as a
wave packet overlap is specific to ideal sources emitting well defined and coherent single 
electron wave packets. But this is not true in general\cite{Dubois:2013-1,Grenier:2013-1}: in full generality,  the proper way to
understand the outcome of an HOM experiment is in terms of overlaps of single electron coherences.
The Wigner representation provides a transparent way to visualize what contributes to the experimental HOM
signals.

\subsubsection{Single electron tomography revisited}

Expressions \eqref{eq:HBT:overlaps} and \eqref{eq:HBT:excess:HOM:exp} enable 
us to understand the recently proposed single electron  tomography protocol\cite{Degio:2010-4} in a very 
simple way. As in the case of optical homodyne tomography\cite{Smithey:1993-1,Lvovsky:2009-1}, the key is to find a
controlled source which, by varying its parameters will enable us to reconstruct the single electron coherence emitted
by the source to be characterized or, equivalently, $\Delta W_1^{(e)}(t,\omega)$.

\medskip

A natural source that spans the whole space of single electron state is an equilibrium reservoir:
depending on its chemical potential, some states will be filled and others will 
be empty. Its Wigner function is constant in 
time and equal to the equilibrium Fermi distribution of the reservoir. 
Let us assume zero temperature for simplicity: 
by varying the chemical potential, the second term in the r.h.s of 
Eq.~\eqref{eq:HBT:overlaps} will select
a small slice in frequency of the unknown Wigner function. 
Eq.~\eqref{eq:HBT:overlaps} expresses that 
the change in current noises when increasing the chemical potential from $\mu_2$
to $\mu_2+d\mu_2$ is proportional to the population of single electron states
within the incoming channel 1 whose energies are between $\mu_2$ and 
$\mu_2+d\mu_2$. If such a state is populated,
the incoming electrons from both sides anti bunch and the noise does not change. 
If such a state is empty, the
electron emitted by the battery into this interval will be partitioned at the beam splitter 
adding a contribution to the
noise. This is the idea of shot noise spectroscopy originally discussed in the 
context of photo assisted 
shot noise\cite{Kozhevnikov:PhD,Shytov:2005-1} and more recently in the 
context of electron quantum optics\cite{Moskalets:2010-1,Degio:2010-4}.

\medskip

To capture the time dependance of the unknown excess Wigner 
function $\Delta W_{1}^{(e)}(t,\omega)$, a time
dependent source is required. As seen in section \ref{sec:Wigner:pulses:small}, 
in the limit of small drive ($eV_0\ll hf$),
the Wigner distribution issued by a reservoir driven by a small sinusoidal voltage 
$V_d(t)=V_0\cos{(2\pi n ft+\phi_0)}$ is given at first order by:
\begin{equation}
\label{eq:HBT:tomography:response-wigner}
\frac{\partial \Delta W_2^{(e)}(t,\omega)}{\partial (eV_0/hf)}=
 -2\cos{(2\pi n ft+\phi_0)}F_{\mu,T_{\mathrm{el}}}(\omega)
\end{equation}
where $F_{\mu,T_{\mathrm{el}}}(\omega)$ is the convolution of the
characteristic function of the interval $[\hbar^{-1}\mu-\pi f,\hbar^{-1}\mu+\pi f]$ with 
$-(4k_BT_{\mathrm{el}}\cosh^2{(\hbar\omega/2k_BT_{\mathrm{el}})})^{-1}$.
Equation \eqref{eq:HBT:tomography:response-wigner} combined to 
Eq.~\eqref{eq:HBT:excess:HOM:exp} at zero frequency
shows that in the Wigner distribution point of view, the response of the low frequency noise 
to a small a.c. drive applied on the incoming channel 2 
contains the information about the Fourier transform in time of the excess Wigner distribution of the source to be characterized.
In practice,  the finite electronic temperature introduces a blurring of the Wigner function along the energy direction over a scale $k_BT_{\mathrm{el}}$.

\medskip

Approaching single electron tomography from the Wigner function 
point of view immediately rises the question of alternative tomography protocols
based on other reference signals. This question is strongly motivated by the
problem of efficiency: single electron tomography based on HOM-like experiments
rely on ultra-high sensitivity current noise measurements which can be quite costly
in terms of acquisition time\cite{Bocquillon:2013-1}. 

\medskip

Our generic tomography protocol\cite{Degio:2010-4} reconstructs the unknown
excess Wigner function $\Delta W^{(e)}_{1}(t,\omega)$ by extracting from noise measurements
its $t$ dependence at fixed $\omega$ as a Fourier series. The various examples discussed in the
present paper suggest that getting an accurate view of these coherences may require heavy sampling
in $\omega$ as well as in the number of harmonics. 
It would thus be highly desirable to design alternative protocols based on a different controlled source
so that a much lighter sampling is required to obtain an accurate picture of the Wigner function. This problem
is reminiscent of the problematic of compressive sensing\cite{Book:Compressed-Sensing,Candes:2006-1}
but here we deal with single electron coherence
which is a ``quantum signal''. The HOM experiment automatically generates the overlap between
two such ``quantum signals''. The practical problem is then to find which 
controlled source could lead to an
accurate approximation of the unknown Wigner function through a minimal number of 
noise measurements in an HOM experiment.

This problem may lead to some interesting theoretical developments but in practice, 
very few well controlled sources are available. The ac and dc drive lead to our generic
tomography protocol. The energy resolved excitations discussed in 
Sec.~\ref{sec:Wigner:sources:capacitor} 
cannot be considered as controlled due to decoherence between the mesoscopic
capacitor and the QPC. Indeed, characterizing decoherence effects on single electron excitations
is precisely one of the main motivation for single electron tomography.
On the contrary, voltage pulses can be considered as controlled
to some extend: high frequency current measurements\cite{Bocquillon:2012-2} 
can be used to control the shape of the pulse arriving onto the QPC, although in a
limited bandwidth. Moreover, in some cases, 
noise measurement can help characterizing purely electronic pulses\cite{Dubois:PhD,Dubois:2013-1}.
Specific voltage pulses may thus provide interesting families of functions for
a reconstructing single electron coherence at minimal cost in some cases. 
However, finding out the appropriate family of excitations requires an a priori knowledge of 
the signal to be measured. Forthcoming
and foreseeable experiments will for sure involve decoherence effects. Therefore, designing alternative
tomography protocols based on HOM experiments requiring less measurements calls for
an in depth modeling of decoherence in the Wigner function formalism. This would go beyond the
scope of the present paper but is of clear interest from this signal processing perspective.

\section{Conclusion}

In this paper, we have introduced the time-frequency representation of the single electron
coherence in quantum Hall edge channels generalizing the quasi-probability distribution 
function introduced in quantum mechanics\cite{Wigner:1932-1} and commonly used in quantum 
optics. We have shown that this real valued function of time
and frequency provides a very convenient way to image single electron coherence in a quantum 
conductor. 
It gives access to both the energy content and real time development of 
excitations emitted by electron sources, shedding new light on electron
coherence and discussing it with a signal processing language. It also 
provides a natural framework to deal with the quantum nature of electron 
excitations, in which quantum interferences effects have a clear signature 
as pronounced non-classical regions where the Wigner function violates 
the bounds required by the Pauli principle and
for interpreting it as a probability distribution. Along the same lines, it leads 
straightforwardly to the emergent semiclassical picture of the quasiparticle dynamics.

Moreover, the celebrated HBT and HOM interferometry experiments have a very simple interpretation in terms
of overlaps of Wigner function. Consequently, Wigner function computations are directly
useful for interpreting the results of these experiments at both the qualitative and quantitative
levels. In fact, our main message is that the Wigner function is a very relevant representation of
single electron coherence because it provides a simple and unified view of many
single and two particle interference effects that have been experimentally demonstrated 
in electron quantum optics
and quantum nano-electronics over the last fifteen 
years\cite{Ji:2003-1,Neder:2007-2,Roulleau:2007-2,Bocquillon:2012-1,Bocquillon:2013-1}.

Experimentally, measuring single electron coherence or equivalently the Wigner function is an important but
difficult challenge for electron quantum optics. However, as discussed in the present paper, we think that
a variety of techniques is now available to recover information on the Wigner function ranging from 
amplitude interferometry\cite{Haack:2011-1,Haack:2012-2} to 
the HOM experiment and its many possible variants among which the recently proposed 
single electron tomography protocol\cite{Degio:2010-4}. 

To conclude, an important and still open question in electron quantum optics is to compute the Wigner
function taking into account the effects of interactions experienced by the excitations emitted
by a single electron source. In the case of pure single electron excitations, the bosonization
technique already used to discuss the problem of quasi-particle relaxation in quantum Hall
edge channels\cite{Degio:2009-1} can be adapted. This leads to a complete unraveling of 
decoherence scenarios of single electron excitations in quantum Hall edge channels which
will be discussed in a future publication.

\acknowledgements{We thank B.~Roussel from ENS Lyon and Q.~Lavigne from Telecom 
St.~Etienne for their precious help in data visualization. We also thank J.M.~Berroir 
and B.~Plaçais from LPA as well as 
Th.~Jonckheere, Th.~Martin, J.~Rech and C.~Wahl from CPT Marseille for useful discussions.
We finally thank P.~Borgnat from ENS Lyon for discussions about 
time/frequency representation in signal processing and compressed sensing.
This work is supported by the ANR grant ''1shot'' (ANR-2010-BLANC-0412).}

\medskip

\appendix
\section{Properties of single electron coherence}
\label{appendix:coherence}

\subsection{Basic properties}

As in photon quantum optics, first order coherences satisfy the hermiticity property:
\begin{subequations}
\label{eq:coherences:hermiticity}
\begin{eqnarray}
\mathcal{G}^{(e)}_{\rho}(x,t;y,t')^* & = & \mathcal{G}^{(e)}_{\rho}(y,t';x,t) \\
\mathcal{G}^{(h)}_{\rho}(x,t;y,t')^* & = & \mathcal{G}^{(h)}_{\rho}(y,t';x,t)\,. 
\end{eqnarray}
\end{subequations}
In a region of free propagation at velocity $v_{F}$, the single electron coherence obeys:
\begin{equation}
\mathcal{G}^{(e)}_{\rho}(x+v_{F}\tau,t+\tau;x'+v_{F}\tau,t'+\tau)
=\mathcal{G}^{(e)}_{\rho}(x,t;x',t')\,.
\end{equation}
When considering coherences at a position $x$ where propagation can be assumed
to be free, Eq.~\ref{eq:SPC:eh} translates into a relation between electron and hole
coherences in the time domain:
\begin{equation}
\mathcal{G}^{(e)}_{\rho,x}(t,t')=v_{F}^{-1}\delta(t-t')-\mathcal{G}^{(h)}_{\rho,x}(t',t)\,.
\end{equation}
Taking the Fourier transform of this equation with respect to $t-t'$ immediately leads to
the relation between electron and hole Wigner functions \eqref{eq:Wigner:e-h-relation}.

\medskip

In order to connect single electron coherence to the fermionic occupation number, we
decompose 
\begin{equation}
\psi(x,t)=\int e^{-i\omega t}c(\omega)\,\frac{d\omega}{\sqrt{2\pi v_F}}
\end{equation}
where $c(\omega)$ and $c^{\dagger}(\omega)$ respectively create and destroy an energy
at energy $\hbar\omega$ with respect to the reference Fermi energy. These operators
satisfy the canonical anti commutation relations: $\{c(\omega),c(\omega')\}=0$ and
$\{c(\omega),c^{\dagger}(\omega')\}=\delta(\omega-\omega')$. 

\subsection{Cauchy-Schwarz inequalities}

Finally, the single electron coherence satisfies a Cauchy-Schwarz inequality
associated with the Hermitian product on the operator space: $(A|B)=\mathrm{Tr}(B\rho A^{\dagger})$. 
Due to the non zero coherence of the Fermi sea, this inequality is most useful
in the frequency domain at zero temperature. 
Using $A=c(\omega_{-})$ and $B=c(\omega_{+})$ and Eq.~\eqref{eq:SPC:frequency-domain},
the Cauchy-Schwarz inequality writes:
\begin{equation}
\label{eq:Cauchy-Schwarz}
\left|\widetilde{\mathcal{G}}^{(e)}_{\rho,x}(\omega_{+},\omega_{-})\right|^{2}
\leq \widetilde{\mathcal{G}}^{(e)}_{\rho,x}(\omega_{+},\omega_{+})\,
\widetilde{\mathcal{G}}^{(e)}_{\rho,x}(\omega_{-},\omega_{-})\,.
\end{equation}
Decomposing $\widetilde{\mathcal{G}}^{(e)}_{\rho,x}$ into a Fermi sea contribution at chemical potential
$\mu=0$ and an excess contribution which we assume to be regular
\begin{equation}
\widetilde{\mathcal{G}}^{(e)}_{\rho,x}(\omega_{+},\omega_{-})=\frac{2\pi}{v_F}\delta(\omega_{+}-\omega_{-})
\Theta(-\omega_{+})+\Delta\widetilde{\mathcal{G}}^{(e)}_{\rho,x}(\omega_{+},\omega_{-})
\end{equation}
and taking $\omega_{+}$ and $\omega_{-}$ positive shows that the Fermi sea contribution vanishes
and leads to the following inequality satisfied over the (e) quadrant:
\begin{equation}
\label{eq:Cauchy-Schwarz:excess}
\left|\Delta \widetilde{\mathcal{G}}^{(e)}_{\rho,x}(\omega_{+},\omega_{-})\right|^{2}\leq
\Delta \widetilde{\mathcal{G}}^{(e)}_{\rho,x}(\omega_{+},\omega_{+})\,
\Delta \widetilde{\mathcal{G}}^{(e)}_{\rho,x}(\omega_{-},\omega_{-})\,.
\end{equation}
Considering the hole coherence and $\omega_{\pm}<0$, the same line of reasoning shows that
the inequality \eqref{eq:Cauchy-Schwarz:excess} is also true on the (h) quadrant.

Finally, let us consider $\omega_{-}<0$ and $\omega_{+}>0$ and assume that the excess 
single electron coherence does not contribute to the electron occupation below the Fermi level
$\Delta\widetilde{\mathcal{G}}^{(e)}_{\rho,x}(\omega_{-},\omega_{-})=0$. Then, the general inequality
\eqref{eq:Cauchy-Schwarz} implies that $\Delta\widetilde{\mathcal{G}}^{(e)}_{\rho,x}(\omega_{+},\omega_{-})$ also
vanishes: there are no electron/hole coherences. The same reasoning can be done on hole coherences and
shows that there are no electron/hole coherences if the source does not contribute to the excess
occupation number of electrons. Finally, the inequality \eqref{eq:Cauchy-Schwarz:excess} also shows
that if a source does not contribute to the excess hole (resp. electron) occupation number, it has vanishing coherence in
the (h) quadrant (reps. (e) quadrant). This proves that, at zero temperature, as soon as a source does not lead to an excess of
electrons (resp. holes), it only contributes to the excess coherence in the (h) (resp. (e)) quadrant. In other words,
it only generates hole (resp. electronic) excitations.

\section{Energy resolved electronic wave packet}
\label{appendix:wavepacket}

It is possible to obtain an analytic formula for the truncated Lorentzian wave  packet 
\eqref{eq:lorentzian} in the time domain. Depending on the sign of $t$, we build a closed
contour that contains the positive real axis and either the positive or negative imaginary axis.
The integral over the imaginary half-axis can then be expressed in terms of the exponential
integral function\cite{Book:GradRyz} $\mathrm{Ei}(x)$. This leads to:
\begin{equation}
\label{eq:lorentzian:time-domain}
\varphi_{e}(t)= \frac{-i\mathcal{N}_{\mathrm{e}}}{\sqrt{v_F\tau_{\mathrm{e}}}}\,
e^{-\gamma_{\mathrm{e}}t-i\omega_{\mathrm{e}}t}\left(
\Theta(t)-\frac{i}{2\pi}\mathrm{Ei}\left[\left(\frac{\gamma_{\mathrm{e}}}{2}
+i\omega_{\mathrm{e}}\right)t\right]\right)\,
\end{equation}
where 
$\mathcal{N}_{\mathrm{e}}=\left(\frac{1}{2}+
\frac{1}{\pi}\arctan{\left(\frac{2\omega_{\mathrm{e}}}{\gamma_{\mathrm{e}}}\right)}\right)^{-1/2}$
is the normalization factor. The exponential integral part is responsible for the non 
vanishing of the wave packet for $t<0$ and for the oscillations of its enveloppe at $t>0$.
These features can be observed on the numerical evaluation of 
excess single electron coherence $\varphi_{e}(\bar{t}+\tau/2)
\varphi_{e}(\bar{t}-\tau/2)^{*}$ depicted on Fig.~\ref{fig:triptique}(c) as well as on
on the current density $v_{F}|\varphi_{e}(t)|^2$ computed from the Wigner
function (see panels (iii) on Fig.~\ref{fig:MZI:Wigner:SES}). 
Note that Eq.~\eqref{eq:lorentzian:time-domain} predicts
a $\log{(t/\tau_{\mathrm{e}})}$ singularity of $\varphi_{e}(t)$ at $t=0$ 
which is not visible on the numerics due to the UV cutoff in numerical integration.


\end{document}